\documentclass{article}

\usepackage{arxiv}

\usepackage{url,hyperref}

\usepackage{import}

\usepackage{subcaption}
\usepackage{float}

\usepackage{pifont}

\usepackage{xcolor} 

\usepackage{tikz}
\usetikzlibrary {shapes.geometric}
\usetikzlibrary{positioning}

\usepackage{paralist}

\usepackage{amsmath} 
\usepackage[noabbrev, nameinlink]{cleveref}

\usepackage{booktabs}

\usepackage{natbib}

\newcommand{\pkg}[1]{\texttt{#1}}
\newcommand{\proglang}[1]{\texttt{#1}}
\newcommand{\code}[1]{\texttt{#1}}

\definecolor{recarrow}{HTML}{d88dbe}

\newcommand{\given}{\,\vert\,}

\definecolor{green}{HTML}{136F63}
\definecolor{red}{HTML}{C97064}
\definecolor{yellow}{HTML}{FC9120}

\newcommand{\cg}{\textcolor{green}{\ding{52}}}
\newcommand{\cs}{\textcolor{red}{\ding{52}}}
\newcommand{\xg}{\textcolor{green}{\ding{54}}}
\newcommand{\xs}{\textcolor{red}{\ding{54}}}
\newcommand{\qm}{\textcolor{yellow}{\textbf ?}}

\newcommand{\classification}[6]{
  \begin{table}[H]
    \centering

    \begin{tabular}{*{6}{c}}
      M & C (B) & R (2) & S & L & I\\
      \hline
      #1 & #2 & #3 & #4 & #5 & #6
    \end{tabular}
  \end{table}
}

\usepackage{soul}
\renewcommand{\hl}[1]{#1}
\renewcommand{\st}[1]{#1}

\usepackage[utf8]{inputenc} 
\usepackage[T1]{fontenc}    

\title{Human Ancestries Simulation and Inference: A Review of Ancestral Recombination Graph-Based Approaches}
\renewcommand{\shorttitle}{ARG Software Review}

\newif\ifuniqueAffiliation
\uniqueAffiliationtrue

\ifuniqueAffiliation
\author{
\href{https://orcid.org/0000-0003-1992-2155}{\includegraphics[scale=0.06]{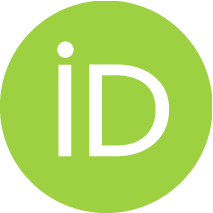}\hspace{1mm}Patrick Fournier} \\
  \\
  Département de Mathématiques\\
  Université du Québec à Montréal\\
  Montréal, Québec, H3C 3P8 \\
  \texttt{pf@patrickfournier.ca} \\
   \And
 Fabrice Larribe \\
  Département de Mathématiques\\
  Université du Québec à Montréal\\
  Montréal, Québec, H3C 3P8 \\
  \texttt{larribe.fabrice@uqam.ca} \\
}

\hypersetup{
pdftitle={Human Ancestries Simulation and Inference: a Review of Ancestral Recombination Graph-Based Approaches},
pdfsubject={q-bio.PE},
pdfauthor={Patrick Fournier, Fabrice Larribe},
pdfkeywords={Coalescent Theory, Ancestral Recombination Graph, ARG Inference, ARG Simulation, Software Review},
}

\begin{document}
\maketitle

\begin{abstract}
 There is little debate about the importance of the ancestral recombination graph in population genetics. An important theoretical tool, the main obstacle to its widespread usage is the computational cost required to match the ever-increasing scale of the data being analyzed. Many of these difficulties have been overcome in the past two decades, which have consequently seen the development of increasingly sophisticated ARG simulation and inference software. Nonetheless, challenges remain, especially in the area of ancestry inference. This paper is a comprehensive review of ARG simulation and inference programs that have emerged in the past three decades to meet the need for scalable and flexible ancestry simulation and inference solutions. It specifically focuses on their performance, usability, and the biological realism of the underlying algorithm, and aims primarily to provide a technical overview of the field for researchers seeking to design and implement their own coalescent-with-recombination algorithm. As a complement to this article, we have compiled links to software, source code and documentation and made them available at \url{https://patrickfournier.ca/publications/arg-software-review/graph}.

\end{abstract}


\section{Introduction}
The ancestral recombination graph (ARG) has been described by some
\cite{Hubisz2020a} as the ``holy grail'' of statistical population genetics.
It is, by all accounts, of great value to the analyst, and while less elusive
than the proverbial Arthurian relic, many challenges have hindered its
widespread adoption. Among these, the heavy computational burden associated with
its simulation and inference from genetic data is probably the most noteworthy.
It is true that many of these difficulties have been overcome in the past
two decades, which have consequently seen the development of increasingly
sophisticated and performant solutions. Nonetheless, challenges remain,
especially in the area of ancestry inference.

This paper reviews ARG simulation and inference software from the early 2000s to the present
day. As such, it is not the first review of its kind. Previous work include:
\begin{itemize}
  \item \citeauthor{CarvajalRodriguez2008}~\cite{CarvajalRodriguez2008}: a
    short comparison of the features of various coalescent-with-recombination
    (CWR) simulators as well as some forward-in-time approaches not based on the
    coalescent;
  \item \citeauthor{Liu2008}~\cite{Liu2008}: similar to
    \cite{CarvajalRodriguez2008} but compares fewer features;
  \item \citeauthor{Arenas2012}~\cite{Arenas2012}: a comparison of CWR,
    forward-time, and phylogenetic simulators, with a focus on evolutionary
    scenarios and mutation models;
  \item \citeauthor{Yuan2012}~\cite{Yuan2012}: a comparison of features,
    computational complexity and execution time of  CWR and forward-time
    simulators, along with the mathematical basis of the underlying models and a
    brief presentation of resampling approaches;
  \item \citeauthor{Hoban2012}~\cite{Hoban2012}: an extensive comparison of
    the features of forward-time and CWR simulators, along with practical
    application considerations;
  \item \citeauthor{Yang2014}~\cite{Yang2014}: a comparison of 5 simulators
    revolving around the simulation of recombination hotspots;
  \item \citeauthor{Brandt2024}~\cite{Brandt2024}: a review of three ARG
    inference software, along with a table summarizing most of the available
    inference programs as of 2023.
  \item \citeauthor{Nielsen2024}~\cite{Nielsen2024}: A short and to the point
    introduction to coalescent theory with a comparison of major inference
    software featuring a small-scale simulation study.
\end{itemize}

In comparison, our evaluation is differentiated by three aspects. First, it is very extensive, if not exhaustive. We review 33 programs in detail and mention 8 others. \hl{We emphasize that our review is limited to programs capable of outputting ARGs. In particular, pieces of software designed to produce exclusively lower-dimensional objects such as genotype matrices and summary statistics are excluded. From that perspective, our work encompasses a broader range of tools than any of the above-listed articles.} Second, it includes both simulation and inference software (see \cref{simulation-inference}). Third, and most importantly, it aims primarily to provide a technical overview of the field for researchers seeking to write their own simulation software. It \st{specifically} focuses on performance, usability, and biological realism of the underlying algorithm. \hl{Specifically, we provide overviews of algorithms, highlight noteworthy innovations, and summarize their relation to other ancestral recombination graph simulation and inference procedures to provide a sense of the current state of the field and its evolution over the last twenty-five years. When appropriate, we describe practical aspects of software packages, such as the application programming interface (API), input and output formats, and additional functionalities.} As such, it is \hl{the only review of this kind and} fundamentally different from \st{the reviews} \hl{those} cited above, which are written with the user rather than the programmer in mind.

\subsection{Vocabulary}
Terminology in coalescent theory is extensive and sometimes confusing. To
minimize misunderstanding, this paper adheres to a strict vocabulary, which is
explained in this section.

\subsubsection{ARG vs Coalescent-With-Recombination}
As noted by \cite{Wu2008,Wong2024}, the term ``ARG'' can be confusing, as it is
not used consistently in the literature. To ensure clarity, we will reserve its
usage to refer to a graph-theoretical object containing information about the
evolutionary history of a sample; this corresponds to the ``data structure''
usage in \cite{Wong2024}. For the other meaning, that of a stochastic process,
we use the term \emph{coalescent-with-recombination} or the acronym CWR.

\subsubsection{Location vs Position}
Recombination events are inherently two-dimensional. They are characterized by
their location on the ARG and on the sequences. We refer to these quantities
as the \emph{location} and the \emph{position} of the event, respectively.
The meaning is preserved when we write that an event is \emph{located} or
\emph{positioned} somewhere.

\subsubsection{Recombination, Chromosomal Crossover and Gene Conversion}\label{recombination}
For the purpose of this paper, ``recombination'' denotes a class of genetic
processes involving the exchange of material between chromosomes. Chromosomal
crossover and gene conversion are both members of this class, with the former
being the main focus of ARG-based approaches. This is not surprising, considering its
biological significance and its relevance for genetic analysis. In fact, the
meaning of the term ``recombination'' is often restricted to ``crossover'' in
much of the literature on coalescent theory.

Chromosomal crossover is the process by which two chromosomes swap portions
of their material. Gene conversion also involves two chromosomes, but is more
unidirectional in comparison; it results in a single chromosome being modified.
A segment of the donor simply replaces the homologous segment in the recipient
chromosome.

\subsubsection{Simulation vs. Inference}\label{simulation-inference}
ARG \emph{simulation} aims to generate an ARG from a set of genetic parameters
such as recombination rate and effective population size. The set of haplotypes
associated with vertices that have no children constitutes a sample of genome
sequences and is often of primary interest. ARG \emph{inference} is the inverse
problem of producing ancestries that are ``consistent'' with a given sample
of haplotypes. Similarities and differences between these two procedures are
discussed in \cref{inferred-vs-simulated}.

\section{Typology}
This section details the broad categories guiding our review.

\subsection{Model vs Heuristic}
Model-based algorithms generate events according to predetermined probability
distributions. These include, without being limited to, coalescences,
chromosomal crossover, mutations, and migration. Their main use is to generate an
ancestry using parameters estimated from genome data. The resulting graphs are
in turn useful for estimating quantities including the time to the most recent
common ancestor (TMRCA) of positions of interest.

On the other hand, heuristic-based algorithms give little to no importance to
probability distributions. A major heuristic is the parsimony of mutation and
recombination events. Algorithms designed accordingly follow the general
idea that ancestries with fewer events are more likely to resemble real-world
genealogies.

Although heuristic algorithms produce events according to broad guiding principles
rather than probability distributions, they are not inherently deterministic.
Finding an evolutionary network minimizing the number of recombinations, for
instance, is NP-hard. For that reason, many ARG inference programs rely on greedy algorithms
which can reach states where the next element must be chosen from a set in
which all elements equally satisfy the heuristic. In such situations, it is not
unusual for the software to choose its next state randomly. The distribution of
that state, however, is expected to be uninformative, as it exists merely as a
convenient way of dealing with nearsightedness rather than fitting data.

The level of performance achieved by parsimonious algorithms often constitutes
their main appeal. Generating ARGs according to a probability distribution is, in itself, a
computationally demanding task that they largely avoid. Even when necessary,
random number generation is generally limited to continuous or discrete uniform
distributions, which are the easiest to sample from. Perhaps more importantly,
parsimony means that the number of generated events itself is likely reduced.
Combined with careful implementation, this makes them difficult to beat on any
time or space-measuring benchmark.

Notwithstanding the intuitive appeal of the various heuristics implemented by
this class of algorithms, model-based ones are the most natural from a statistical
standpoint. At the end of the day, genetic data is data, and a major use for
ARG simulation and inference in an analysis pipeline is to produce genealogies from a user-specified
distribution.

In other contexts, the goal is to infer the ancestry of a sample of haplotypes.
The choice between model-based and heuristic-based solutions is often dictated by
methodological considerations and the complexity of the task. One's preference
for the statistical rigor of a mathematical model of population evolution
might be tempered by the reality that procucing a genealogy for a large set of
individuals with chromosome-scale numbers of genetic markers might be impossible
with the computational resources at hand. There is also a case to be made for
preferring heuristics when it comes to this kind of punctual estimation. That
being said, CWR-based programs can also be used for purposes requiring the generation
of multiple ancestries, such as interval estimation or numerical integration.
Since both require generating multiple ancestries, heuristic-based approaches that
invariably produce the same output given a particular input are not an option.
Even those that introduce randomness are probably not ideal choices, as the
quality of the genealogies is generally a primary concern for those tasks. This is,
of course, unfortunate, as their number comes second in importance.

\subsection{Supported Events}
Perhaps the most common way of reducing the computational burden is
to reduce the space of admissible events. Early programs were more or
less direct implementations of mathematical developments, a list of which can
be found in \cite{Lewanski2024}. Before long, algorithms avoiding the simulation
of non-informative events were developed for performance's sake. Starting in
the mid-2000s, simulators prohibiting some informative events as well were
developed. While this subclass of applications can only approximate the coalescent with recombination, they can achieve remarkable performance
that is often orders of magnitude superior to their exact counterpart.

\subsubsection{Coalescence events}\label{coalescence-events}
It is customary to categorize coalescence events based on the ancestrality structure of the generic material associated with the coalescing sequences. We distinguish between material inherited by the sample and material that is not. The first kind is called \emph{ancestral} material, while material of the second kind is called \emph{non-ancestral}. Given two sequences, we classify their coalescence as a \emph{type A} event if some of their ancestral material overlaps. Otherwise, the event is deemed to be of \emph{type B}. Regardless of its type, the occurrence of any coalescence event has the potential to affect the composition of the sample. Consequently, the choice of disregarding a particular type is made primarily, if not solely, for performance reasons.

\begin{figure*} 
  \centering
  \begin{subfigure}[t]{0.45\textwidth}
    \centering
    \includegraphics[width = \columnwidth]{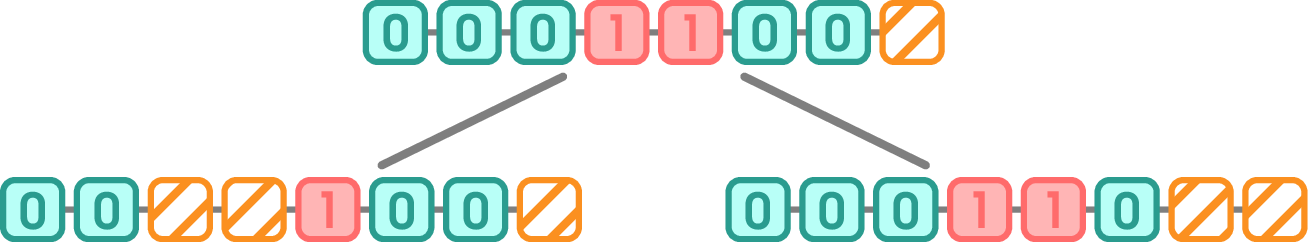}
    \caption{Type A}\label{fig:coalescence-a}
  \end{subfigure}
  \hfill
  \begin{subfigure}[t]{0.45\textwidth}
    \centering
    \includegraphics[width = \columnwidth]{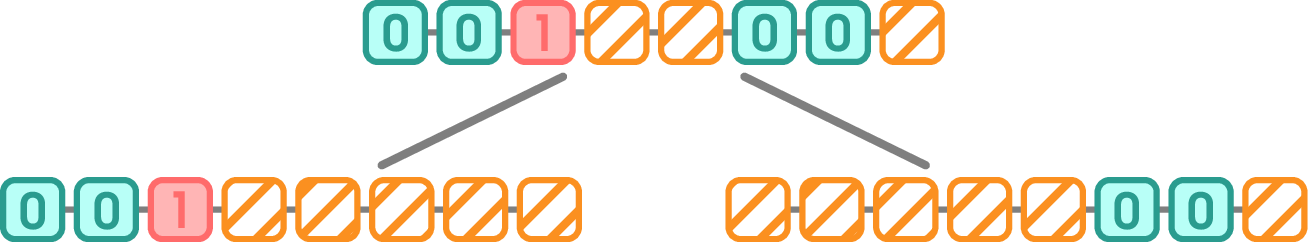}
    \caption{Type B}\label{fig:coalescence-b}
  \end{subfigure}
  \caption{Examples of coalescence events.}\label{fig:coalescence}
\end{figure*}

Disregarding type A events has never been done in practice. There are no known performance gains to be achieved in this way. On the other hand, denying type B events led to the now famous \emph{sequential Markov coalescent} (SMC). This algorithm builds on a well-known article by Wiuf and Hein, and overcomes one of its limitations. In \cite{Wiuf1999}, they conceptualize the ARG in a novel way as a coalescent tree to which edges and vertices representing recombination and recoalescence events are added iteratively. They call their construct the Spatial Algorithm Graph (SAG) to contrast it with earlier \emph{temporal} algorithms that build graphs backward in time from a sample. From then on, the adjective ``spatial'' has been used to refer to any ARG construction algorithm that moves along sequences rather than time.

This shift in perspective, however, doesn't provide any fundamental increase in performance. The main reason is that to correctly follow the CWR distribution, recombination locations must be modeled as a Poisson process on the complete graph at each step of the algorithm. Selecting the next event is consequently more demanding than the last, compounding the computational burden until the end of the sequences is reached. \citeauthor{McVean2005}~\cite{McVean2005} studied a slight modification of the SAG where the location of a recombination event is constrained to the subgraph induced by vertices ancestral for a position. This induced subgraph is a full binary tree and is called the \emph{marginal tree} of the position. Their algorithm is Markovian insofar as, at any given step, the recombination event conditional on the current marginal tree is independent of the remainder of the graph. This slight modification has a tremendous impact on the algorithm's performance, as the compounding effect mentioned earlier no longer occurs. From a computational complexity standpoint, a Markovian model of the dependence structure of sequence segments opens the door to linear-time algorithms with respect to sequence length. Of course, as noted by Wiuf and Hein, the distribution of ancestries produced in this way only approximates the ``true'' distribution of the ancestral recombination graph. In fact, as noted by Cardin and McVean, their scheme is equivalent to disallowing type B events. Nevertheless, the SMC algorithm, which exploits this approximation by operating on sequences of trees rather than graphs, is arguably the best-known result from the work of Cardin and McVean. More flexible SMC-type approximations allowing events to be located on the subgraph induced by vertices ancestral for an interval rather than a singular position were developed later.

Theoretical results concerning the effects of the SMC approximation on parameter estimation are generally difficult to obtain, but some are available. \cite{Wilton2015} use a two-locus model to show equality in distribution between the pairwise joint coalescence time at two recombination sites under ARG and SMC' \cite{Marjoram2006}, a slightly modified version of the SMC algorithm, along with some more general numerical results. \cite{Chen2014} shows equality in distribution between ancestries simulated by what they call a \emph{typical back-in-time algorithm} (\pkg{ms} with minor modifications), and their own generalized version of the SMC called SC. Their algorithm offers a spectrum of approximations ranging from the SMC, where events' locations are restricted to the current marginal tree, to no restriction at all. A brief overview of the main differences between SMC and SMC' is available in \cref{fastcoal}.

Simulation studies about SMC's accuracy are more readily available, starting with the original paper \cite{McVean2005} and Cardin's own PhD thesis \cite{Cardin2007}. \cite{Eriksson2009} shows that the SMC' approximation is generally accurate under various demographic structures regarding the pairwise correlation between TMRCA and the linkage probability of two loci. \cite{Hobolth2014}, among other things, studies the accuracy of recombination and coalescence rate estimation under SMC and SMC' approximations with respect to their own ``natural'' Markovian approximation of the CWR. They conclude that SMC' is overall quite accurate. One important caveat is that estimations of the recombination rate might be inflated when it is an order of magnitude greater than the coalescence rate. SMC, however, systematically underestimates the recombination rate. \cite{Shchur2022} shows that the total variation distance between the pairwise joint distribution of the TMRCA for SMC' and the full CWR model is non-negligible under a demographic model of two populations with symmetric migration.

\subsubsection{Recombination events}
Similarly, recombination events can be classified with respect to their relation to the ancestral material of the sequence on which they occur. We distinguish between five types of events. The presence of ancestral material on both sides of the recombination position is the defining characteristic of the first two types. Such an event is of type 1 if it occurs within ancestral material, and of type 2 otherwise. Types 3 and 4 events correspond to cases in which ancestral material is present only to the left or the right side of the position, respectively. An event positioned on a sequence devoid of ancestral material is of type 5.

Unlike coalescence events, some types of recombination events have no bearing on the sample. The outcomes of type 3, 4, and 5 events are exclusively either non-ancestral sequences that can be safely disregarded or clones of existing haplotypes. These events are targeted for optimization by all reviewed software. Types 1 and 2 events, on the other hand, have the potential to be consequential for the sample and are desirable from the point of view of biological realism. Exact algorithms are largely indifferent to the type of recombination event; both are equally easy to simulate from a computational standpoint. This is not as straightforward for approximate spatial algorithms. Consider that from a spatial point of view, any recombination event located in the marginal tree of its position must almost by definition be of the first type since every edge of the tree is ancestral for that position. This is to be contrasted with the fact that an event of the second type cannot be located in the marginal tree associated with its position. Marginal tree-centric algorithms -- such as the original SMC or variations like SMC' -- are consequently unable to generate events of the second type. This is almost self-evident for SMC and SMC', as the kind of trapped non-ancestral material required for the occurrence of type 2 recombination can only be created through type B coalescence, which they deliberately omit. Trapped material and type 2 events may \cite{Wang2014, Mahmoudi2022} or may not \cite{Rasmussen2014} have a significant impact on the sample.

We summarize this section as follows: when optimizing model-based algorithms, there is a tradeoff to be made between realism and performance. The main optimization target concerns trapped non-ancestral material, which is a byproduct of type B coalescence events. Denying such events leads to a Markovian correlation structure and a linear-time algorithm. However, ancestry produced this way also lacks type 2 recombination events. Interestingly, it is possible to generate type B coalescence while excluding type 2 recombination events (see \cref{sec:macs}). Finally, we do not want to give the impression that approximating the CWR is the only possible avenue for optimizing algorithms for performance. While it may be the case that the correlation structure of sequence segments is not Markovian, precluding the existence of linear-time exact algorithms, there is more to the story than asymptotic observations, since sequences of genetic markers are by their very nature discrete and finite objects. While it is certainly of theoretical interest to reduce the computational complexity of ARG simulation and inference, the sole practical requirement (disregarding restrictions related to data storage) is to bring the duration of simulations below a determined threshold for a given sample. These two objectives are of course related, but the importance of efficient implementation in reducing the time necessary for actual computations cannot be overstated (see \cref{sec:msprime}).

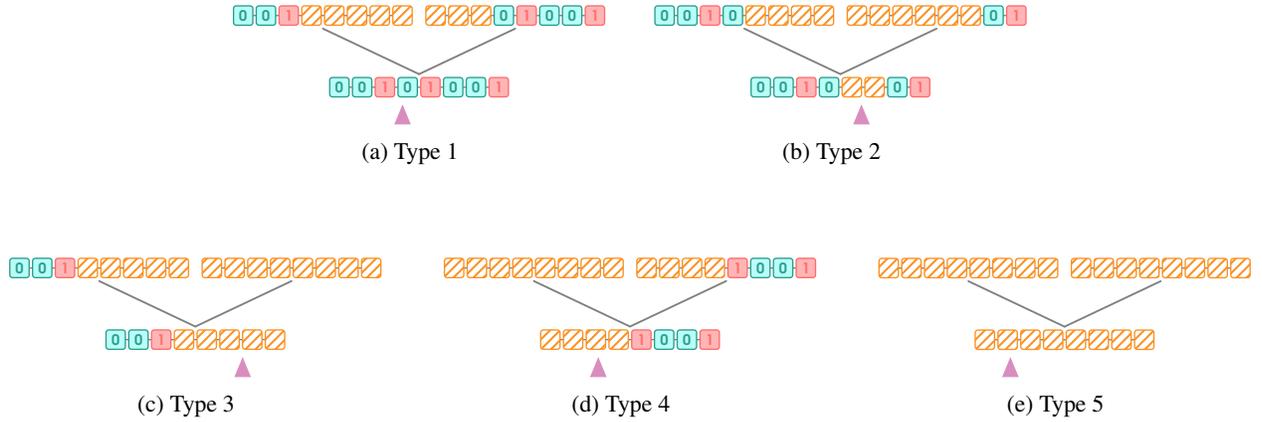
\begin{figure*} 
  \centering
  \begin{subfigure}[t]{0.3\textwidth}
    \centering
    \begin{tikzpicture}
      \node (figure) {
        \includegraphics[width = \columnwidth]{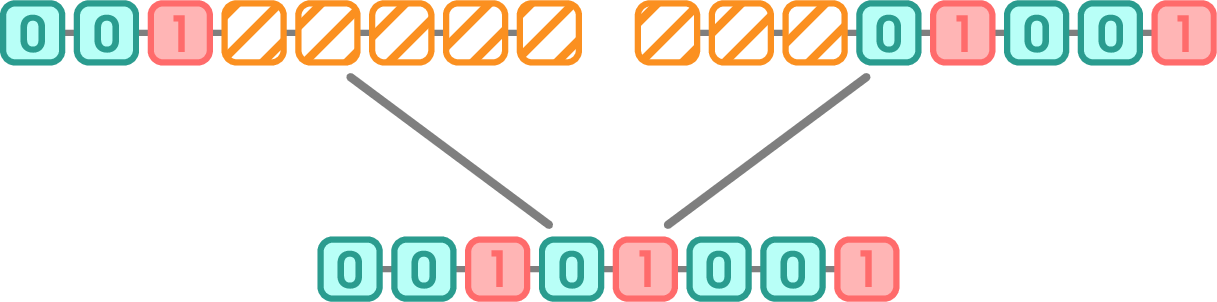}
      };
      \node[isosceles triangle,
            draw = recarrow,
            below = of figure,
            inner sep = 0pt,
            rotate = 90,
            fill = recarrow,
            minimum size = 2mm] at (-0.3, 0.1) {};
      \end{tikzpicture}
    \caption{Type 1}\label{fig:recombinations-1}
  \end{subfigure}
  \hspace{3ex}
  \begin{subfigure}[t]{0.3\textwidth}
    \centering
    \begin{tikzpicture}
      \node (figure) {
        \includegraphics[width = \columnwidth]{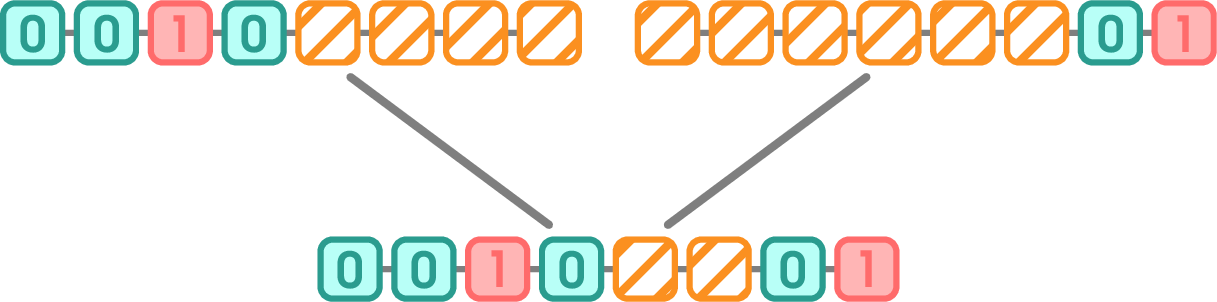}
      };
      \node[isosceles triangle,
            draw = recarrow,
            below = of figure,
            inner sep = 0pt,
            rotate = 90,
            fill = recarrow,
            minimum size = 2mm] at (0.2, 0.1) {};
      \end{tikzpicture}
    \caption{Type 2}\label{fig:recombinations-2}
  \end{subfigure}

  \vspace{3em}
  \begin{subfigure}[t]{0.3\textwidth}
    \centering
    \begin{tikzpicture}
      \node (figure) {
        \includegraphics[width = \columnwidth]{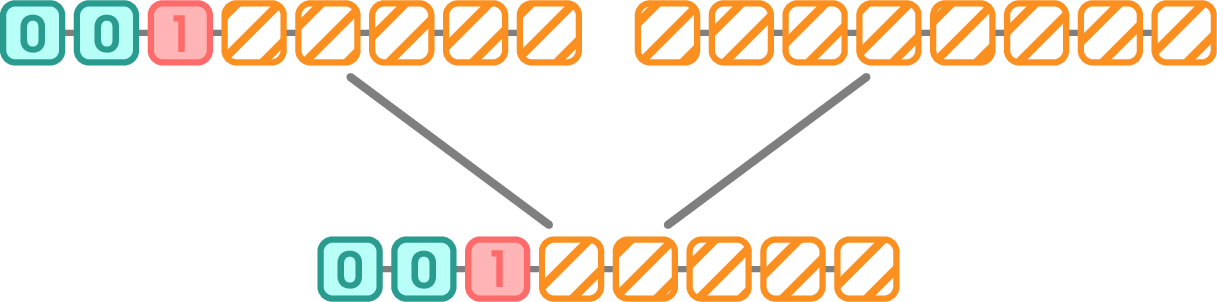}
      };
      \node[isosceles triangle,
            draw = recarrow,
            below = of figure,
            inner sep = 0pt,
            rotate = 90,
            fill = recarrow,
            minimum size = 2mm] at (0.55, 0.1) {};
      \end{tikzpicture}
    \caption{Type 3}\label{fig:recombinations-3}
  \end{subfigure}
  \hfill
  \begin{subfigure}[t]{0.3\textwidth}
    \centering
    \begin{tikzpicture}
      \node (figure) {
        \includegraphics[width = \columnwidth]{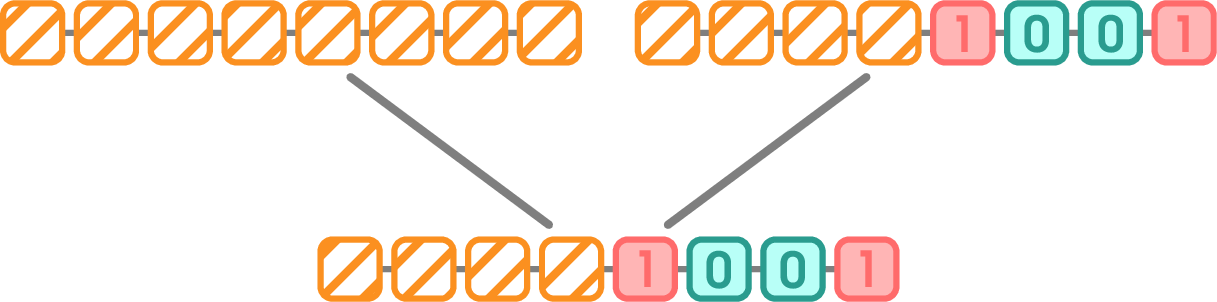}
      };
      \node[isosceles triangle,
            draw = recarrow,
            below = of figure,
            inner sep = 0pt,
            rotate = 90,
            fill = recarrow,
            minimum size = 2mm] at (-0.5, 0.1) {};
      \end{tikzpicture}
    \caption{Type 4}\label{fig:recombinations-4}
  \end{subfigure}
  \hfill
  \begin{subfigure}[t]{0.3\textwidth}
    \centering
    \begin{tikzpicture}
      \node (figure) {
        \includegraphics[width = \columnwidth]{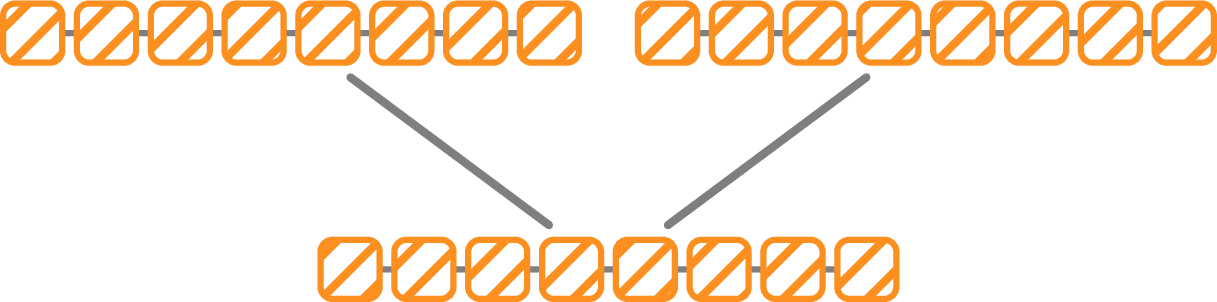}
      };
      \node[isosceles triangle,
            draw = recarrow,
            below = of figure,
            inner sep = 0pt,
            rotate = 90,
            fill = recarrow,
            minimum size = 2mm] at (-0.8, 0.1) {};
      \end{tikzpicture}
    \caption{Type 5}\label{fig:recombinations-5}
  \end{subfigure}

  \caption{Examples of recombination events.}\label{fig:recombinations}
\end{figure*}

\subsection{Inferred vs Simulated}\label{inferred-vs-simulated}
Although these adjectives have been used in the literature for quite some time, they have not been formally defined, to the best of our knowledge. Consequently, we begin this section with the following tentative definition: \emph{ARGs simulation} aim to produce one or multiple ancestries according to a specific genetic model such as Wright-Fisher (WF) or CWR. \emph{ARGs inference} aim to produce one or multiple ancestries that may have given rise, according to some criterion, to a sample of haplotypes.

In practice, parameters for ARG simulation are often estimated from a sample of the target population. These might include, but are not limited to, values such as recombination rate, mutation rate, effective population size, and demographic model. The output can be used to estimate various quantities of interest. Such an approach might seem overly complicated at first glance, but it is very natural when the quantity of interest is ancestry-related. A classical example is the TMRCA of a chromosomal segment.

Inputs for ARG inference software, especially model-based ones, generally include parameter estimates as well. All of them require an initial set of haplotypes to construct ancestries. Although they may aim to produce genealogies distributed according to a specific genetic model, at least approximately, this goal is secondary to consistency with the input haplotypes. The most common notion of consistency, at least for software focussing on human genealogies, stems from \citeauthor{Kimura1969}'s \emph{infinite-site model} \cite{Kimura1969} (ISM) for mutations. The basic idea is that, as long as the locations of the mutations are supported on the continuum, the probability of more than one mutation event happening at the same site is vanishingly small. From this point of view, consistency can be informally defined as follows: an ancestry is consistent with a sample if there is a way to locate mutations on its edges that give rise to the set of input sequences while mutating each locus at most once. The ISM might appear as a strong assumption, but it may be plausible when working at the nucleotide level, as noted by \citeauthor{Kimura1969}. Considering that the human genome is characterized by a relatively large number of nucleotides and a relatively low mutation rate, it is arguably a reasonable model for nucleotide-level markers such as SNPs.

From this point of view, it may seem as if model-based approaches rely on a minimization heuristic for the number of mutation events, much like their parsimony-based counterparts. While there might well be some truth in that statement, we would like to point out that the infinite sites model is a model, not a heuristic. It simply happens to entail parsimony of mutation events. Inference software based on a different set of assumptions are possible, at least in theory. The schism between model and parsimony occurs at the level of recombination events.

\subsection{Programming Language}
The programming language used to implement an algorithm may be of interest for certain use cases. One might, for instance, desire to modify a software to suit their needs, a task that is easier done in a familiar language. The language a software is coded in can also provide some hints about its characteristics. It is reasonable to expect a \proglang{C} program to be more efficient but also less user-friendly than a \proglang{Python} one, for instance.

It is becoming more common for software to be written in multiple languages. A given language is often designed to prioritize one or at most a few aspects of software development. Understandably, programmers want the best of every world, a desire that has become increasingly easy to satisfy. With that in mind, we will classify programs according to the language used to implement the actual algorithm. This decision is natural since, among all the concerns addressed by programming languages, computational performance usually comes first. Ease of use, addressed by the next criterion, is next on the list.

\subsection{Interface}
The interface of a software is to be understood here in a broad sense. It includes any way a user or another software might interact with the algorithm's implementation. For example, we consider a program for which there exist \proglang{Rust} bindings as having a \proglang{Rust} interface. Another example is software implementing an algorithm in \proglang{C} while providing a \proglang{Python} application programming interface (API). We would classify such an application as being coded in \proglang{C} and providing a \proglang{Python} interface.

In order for an interface to be taken into account, two conditions must be met:
\begin{inparaenum}[(1)]
\item it must be documented and
\item a \emph{significant} part of the program must be exposed through it.
\end{inparaenum}

An interface does not need to be a programming language. A command-line interface (CLI), for instance, might be a valid interface.

\section{Review of Software}
Programs are grouped by family where appropriate and then listed in chronological order. Properties of the software are summarized in short tables. Column titles have the following meaning:
\begin{itemize}
  \item M: model-based (vs. heuristic-based);
  \item C (B): supports type B coalescence events;
  \item R (2): supports type 2 recombination events;
  \item S: simulated (vs. inferred);
  \item L: programming language;
  \item I: interface.
\end{itemize}
When a property is binary (e.g. M), a checkmark (\cg or \cs) or a cross mark (\xg or \xs) indicates the presence or absence of the property. A green/red checkmark indicate desirable/undesirable features. The opposite is true for crosses. Only the principal interface of programs with many interfaces is listed. An asterisk ``*'' indicates that other interfaces exist and are discussed in the associated subsection. Unknown status is indicated by a yellow question mark (\qm).

\subsection{\pkg{ms} Family}
Simulators in this family implement algorithms based on Hudson's original work \cite{Hudson1983, Hudson1991}. They are model-based temporal simulators able to generate all types of informative coalescence and recombination events. They generally provide a high degree of accuracy with respect to the CWR.

\subsubsection{\pkg{ms} \cite{Hudson2002}}\label{ms}
\classification{\cg}{\cg}{\cg}{\cg}{\proglang{C}}{CLI}

\pkg{ms} is widely regarded as the reference CWR implementation against which others' statistical performances are benchmarked. To avoid confusion, we reserve the name \pkg{ms} for the simulator itself and refer to the algorithm it implements as ``\emph{Hudson's algorithm}''.

Every informative coalescence and recombination event is supported, and none of the non-informative ones are ever simulated. In addition, \pkg{ms} supports multiple extra features:
\begin{enumerate}
\item gene conversion;
\item demographic events;
\item migration events;
\item inclusion of ancient DNA.
\end{enumerate}
Gene conversion functionalities were added in a 2001 update. When requested by the user, events are simulated according to \cite{Wiuf2000}, a model based on the Holliday Junction \cite{Holliday1964}. Simulations can be carried out with any of the four possible combinations of chromosomal crossover/gene conversion events. Migration events follow an island model. An arbitrary number of subpopulations with variable sample sizes can be specified. It is even possible to introduce population splits at predetermined times. The migration matrix is entirely customizable by the user and can even vary as a function of time. Similarly, growth rates and population sizes can vary dynamically. Ancient chromosomes can be placed anywhere back in time and in any subpopulation.

All in all, the flexibility of \pkg{ms}, brought about by its extensive set of features, is impressive, especially at the time of its release. From a computational performance perspective, \pkg{ms} is generally regarded as efficient. This claim is largely based on the types of events it supports, which are limited to informative ones. Of course, its performance is not on par with programs implementing approximations to the CWR, but such a comparison would be unfair. That being said, \pkg{ms}'s performance is not sufficient to meet the requirements of most contemporary genetic analyses, which are often conducted with large samples and on a genome scale. For instance, \cite{Kelleher2016} exclude \pkg{ms} from their ``performance analysis'' section. The time to complete simulations in the most demanding scenarios is much longer than for more recent benchmarked simulators.

Hudson's algorithm is implemented as a loop that continues as long as the sample's MRCA has not been reached; it is the archetypical backward-in-time simulator. At each iteration, a type of event (coalescence, recombination, mutation, or migration) is generated. The current graph is then altered in accordance with the outcome. This description is rather high-level, as the program itself doesn't store the ARG itself but rather the corresponding sequence of marginal trees.

Conveniently, \pkg{ms} can output this sequence, each tree encoded in Newick format, upon user's request. As noted by Hudson, this comes in handy when interaction with other tools is desired. On the other hand, as we will see in \cref{sec:msprime}, \pkg{ms}'s data structure is also its Achilles heel; it simply doesn't scale, neither with sample size nor sequence length. Interestingly, parallelized versions have been developed (see \cref{sec:ms-other}).

\subsubsection{\pkg{msms} \cite{Ewing2010}}
\classification{\cg}{\cg}{\cg}{\cg}{\proglang{Java}}{CLI}

\pkg{msms} aims to improve on \pkg{ms} by adding selection capabilities. It is nonetheless able to construct neutral ancestries (i.e., without selection) and is equivalent to \pkg{ms} when used in that way. Moreover, their command-line interfaces and text outputs are identical. All of this makes \pkg{msms} a drop-in replacement for \pkg{ms}. The authors claim that its computational performance is generally on par with that of \pkg{ms}. However, reports suggest that it outperforms the original, especially in scenarios involving high recombination rates and/or large sample sizes \cite{Kelleher2016}.

If requested by the user, selection is integrated into the CWR model by simulating ancestries conditional on the major allele's frequency (MAF). MAF is computed prior to the actual ancestry simulation in a forward pass involving a structured Wright-Fisher model with selection. In the case of a spatially structured population, fitness values can be specified on a per-deme basis.

Although \pkg{msms} implements Hudson's algorithm, it differs substantially from \pkg{ms} internally. While both programs' interfaces and output are identical, \pkg{msms} is actually implemented in \proglang{Java} rather than \proglang{C}. As it turns out, performance is not negatively impacted by this choice. In addition, since \proglang{Java} runs on a virtual machine, \pkg{msms}' portability is superior to that of its predecessor, a significant advantage for reproducible research and open science. Moreover, since \proglang{Java} is a higher-level language than \proglang{C}, \pkg{msms} code is arguably easier to understand and modify.

\subsubsection{\pkg{cosi2} \cite{Shlyakhter2014}}
\classification{\cg}{\cg}{\cg}{\cg}{\proglang{C++}}{text}

Published over a decade after \pkg{ms}, \pkg{cosi2} innovates in its implementation of Hudson's algorithm. At the time of its release, simulating long sequences was exclusively the domain of software implementing approximations to the CWR, all of which lacked selection capabilities. Furthermore, no simulator, approximate or exact, had the ability to generate both selection and gene conversion events. An efficient implementation of Hudson's algorithm would allow simulation of these complex scenarios, which is precisely what \pkg{cosi2} aims to achieve. The software's main innovation compared to the original algorithm is its clever exploitation of the backward-in-time CWR Markovian property. At each step of the simulation, the next state can be choosen without knowledge of anything but the current one. Consequently, \pkg{cosi2} discards unnecessary information after each transition. Of course, just because some information can be disregarded at some point in the simulation does not mean it is irrelevant to the final result. It turns out that the complete ARG can be reconstructed by recording, at every step and for every node $n$, the function that maps an interval $x$ to the set of sequences in the sample that inherit material in $x$ from $n$. This can be done efficiently using skip lists. Gains in terms of execution time are substantial, even more so for the approximate version of \pkg{cosi2}, which restricts coalescence events to type A.

Simulation parameters are set in a text file using a custom format. Output is compatible with \pkg{ms}'.

\subsubsection{\pkg{discoal} \cite{Kern2016}}\label{sec:discoal}
\classification{\cg}{\cg}{\cg}{\cg}{\proglang{C}}{CLI}

The main feature of \pkg{discoal} is its support for fixation of neutral alleles caused by linkage with a beneficial mutation, a phenomenon known as \emph{selective sweep}. Both hard (polymorphism elimination) and soft (polymorphism reduction) events are supported (see \cite{Pennings2006} for a detailed and more nuanced presentation). Other features include recurrent mutation, gene conversion, fixation of arbitrary mutations and support for complex demographic scenarios.

\pkg{discoal} is the simulator used by \pkg{selscan} \cite{Szpiech2024}.

\subsubsection{\pkg{msprime} \cite{Baumdicker2021}}\label{sec:msprime}
\classification{\cg}{\cg}{\cg}{\cg}{\proglang{C}}{\proglang{Python}*}

It is hard to overstate the importance of \pkg{msprime} in the current landscape of CWR simulators. Before its publication, it was largely accepted that large-scale simulations were only possible through approximations. \pkg{msprime} demonstrated that a careful implementation of Hudson's algorithm could, for all intents and purposes, outperform sequential approximations even on genome-scale datasets.

\pkg{msprime}, as its name suggests, aspires to be an improved version of \pkg{ms}. Given the consensus around Hudson's program's efficiency, it might seem almost impossible to achieve significant improvements. While this may well be true from a computational complexity standpoint, it may still be possible to realize practical gains by optimizing the underlying data structure. Ancestries produced by \pkg{ms} are stored as sequences of marginal trees, which is a natural choice given the nature of the data. This design is nonetheless inefficient: each recombination event requires the creation of a new marginal tree with a number of vertices and edges on the order of the sample size. The overhead is arguably negligible for modest sample sizes and recombination rates. However, on the genome scale, as evidenced by \pkg{ms}' poor performance in that context, the spatial and temporal demands prove to be a significant limitation.

\pkg{msprime}'s solution exploits the high degree of correlation between nearby marginal trees. The first key insight is that two adjacent trees are nearly identical. Given the left tree, little information is necessary to recover the right one. Earlier versions of \pkg{msprime} encoded this information in the form of a sequence of \emph{coalescence records}, dubbed a \emph{Tree Sequence} \cite{Kelleher2016}. Each record tracks a coalescence event, which is defined as the merging of two segments with ancestral material in a common interval into a single segment. \emph{Common ancestor events} are distinct from \emph{coalescence events}, with the former events including type A and B events and the latter being reserved for type A. In any case, let $c = (a, b)$ be a coalescing pair, $w$ the segment they coalesce into, $t$ the latitude of $w$, and $[l, r)$ the shared ancestral interval. The corresponding coalescence record is the 5-tuple $(l, r, w, c, t)$. Note that the merging of two segments with multiple disjoint common ancestral intervals generates multiple coalescence records, one for each shared interval. Hence, from a topological perspective, a Tree Sequence is a multigraph. The second insight is as follows: if a sample has reached its MRCA for an interval $[l, r)$ at time $t$, there is no fundamental need to keep track of the associated marginal tree before $t$, as it cannot possibly contain information about $[l, r)$ relevant to the sample; all relevant information is contained in the records preceding the one corresponding to the MRCA of $[l, r)$. Consequently, there is no need to simulate the complete ARG to obtain a sample's ancestry; it is sufficient to generate a valid Tree Sequence. This can be achieved without tracking ``partially built trees'', only by maintaining a set containing ancestral segments. This realization is the main source of performance gain of \pkg{msprime} over \pkg{ms}, as storing and managing these trees is a major source of overhead for the latter. Note that this formulation of Hudson's algorithm allows for other optimizations as well, such as segment defragmentation with adjacent intervals. As of version 1.0 \cite{Baumdicker2021}, the Tree Sequence data structure is generalized, primarily to support supplementary information such as mutations. Nevertheless, the main ideas underlying \pkg{msprime}'s performance gains remain unchanged.

Relying on Tree Sequences renders random access to marginal trees impossible. This is perhaps the main drawback of \pkg{msprime}'s choice of data structure. This is arguably not a major concern, though, as many common operations only require \emph{sequential} access, which can be achieved efficiently via \cite{Baumdicker2021}'s algorithm T.

\pkg{Msprime} is distributed as a \proglang{Python} package and can be installed via \pkg{Pip} or \pkg{Conda}. This allows for seamless integration into a \proglang{Python} workflow, which is a welcome convenience given the widespread adoption of this language in bioinformatics and data science. Online documentation also mentions the availability of a container image on Dockerhub, streamlining execution via Docker, Podman, or even Kubernetes. However, as of March 2025, the image lagged behind \pkg{msprime}'s last release by over four years. Simulations can also be run through a command-line interface. Two applications are provided for this purpose. The first, \pkg{msp}, provides full access to \pkg{msprime}'s functionalities. The second, \pkg{mspms}, implements \pkg{ms}' interface and is intended as a drop-in replacement for Hudson's program.

\pkg{Msprime}'s output is a \pkg{tskit} \cite{Kelleher2018} object of type \code{treeSequence}. This allows for considerable flexibility. The \pkg{tskit} ecosystem is vast and offers a wide range of quality software. Many analyses can be conducted within its bounds. Should integration with other languages be needed, APIs are available for \proglang{Python}, \proglang{C}, and \proglang{Rust}. If integration with \proglang{R} is desired, \pkg{tskit}'s authors recommend interfacing through \pkg{reticulate} \cite{Ushey2025}. Finally, \pkg{tskit} can output Tree Sequences as plain text files.

Even though \pkg{msprime} is first and foremost an implementation of Hudson's algorithm, other models are also available. A discrete backwards-in-time Wright-Fisher model \cite{Nelson2020}, which can be thought of as a discrete-time CWR simulator, is provided. Its inclusion aims to address biases caused by long-range correlation emphasized in ancestries constructed from large samples of long sequences. Hybrid schemes where recent history is simulated according to the discrete model and the remainder under classical CWR are possible. SMC and SMC' approximations are implemented via a simple rejection sampling. Although the distributions are correct, the underlying algorithm remains Hudson's and doesn't take advantage of optimization opportunities offered by Markov approximations. These are clearly intended to facilitate quick comparisons and experimentation, not as ``production grade'' solutions. Models including selective sweep events are supported via an implementation of \pkg{discoal}'s algorithm (see \cref{sec:discoal}). Moving beyond the standard CWR, backwards-in-time pedigree, beta coalescent, and Dirac coalescent models are also available.

\subsubsection{Other Programs}\label{sec:ms-other}
\begin{itemize}
\item \pkg{msHOT} \cite{Hellenthal2006}: Allows for user-specified recombination hotspots. In Hudson's original algorithm, recombination breakpoints are generated from a homogeneous point process. \pkg{msHOT} gives users the ability to vary the process' intensity along sequences.
\item \pkg{mbs} \cite{Teshima2009}: Flexible simulation of sophisticated demographic scenarios for biallelic sequences.
\item \pkg{msPar} \cite{Montemuino2014}: Parallel implementation of \pkg{ms} via MPI.
\item \pkg{msParSm} \cite{Montemuino2016}: Similar to \pkg{msPar} but includes various optimizations. \pkg{msParSm} is able to construct genealogies for ``long'' sequences under a high recombination rate.
\end{itemize}

\subsection{\pkg{SIMCOAL} Family}
This family comprises two simulators developed primarily by Laurent Excoffier from 2000 onwards. In addition to simulation capabilities, these programs can output data in a format suitable for \pkg{ARLEQUIN} \cite{Excoffier2010}, allowing for the computation of summary statistics and further analysis.

\subsubsection{\pkg{SIMCOAL2} \cite{Laval2004}}
\classification{\cg}{\cg}{\cg}{\cg}{\proglang{C++}}{CLI}

Up until version 2.0, \pkg{SIMCOAL} did not support recombination events \cite{Excoffier2000}. It was essentially a coalescent simulator with broad support for demographic events. Unlike many similar programs, it did not rely on a continuous time approximation, instead using an admittedly slower discrete generation approach. While detrimental to performance, discrete time facilitated the implementation of support for multiple events per generation. Version 1 of \pkg{SIMCOAL} had the ability to generate multiple demographic events per generation, a capacity extended to coalescent and recombination events in version 2. \cite{Laval2004} argues for the necessity of multiple coalescence events by generation in a discrete framework on the basis of the potential for a rapidly increasing number of lineages under the effect of recombination. From that point of view, \pkg{SIMCOAL2} is well-suited for samples whose size is not significantly below the effective population size.

Despite the article's lack of details about the algorithm, the source code reveals that the first two types of recombination events are supported exclusively; the function \code{TNode::recombinationPosition} is particularly revealing in that respect. Similarly, \code{TDeme::singleCoalescentEvent} and \code{TDeme::multipleCoalescentEvents} indicate that both types of coalescence events are supported.

\pkg{SIMCOAL} outputs both summary statistics about the sample's ancestry and the ancestry itself. The statistics are contained in an \code{ARLEQUIN}-formatted file. The ancestry is available in the form of a sequence of \code{NEXUS} encoded marginal trees in a separate file.

Ancient DNA is supported via an extension. \pkg{Serial SimCoal} \cite{Anderson2004} adds support for the so-called serial coalescent model. In addition to simulation facilities, it includes tools for statistical analysis of generated data.

\subsubsection{\pkg{fastsimcoal} \cite{Excoffier2011}}
\classification{\cg}{\xs}{\xs}{\cg}{\proglang{C++}}{CLI}

\pkg{fastsimcoal} is a complete rewrite of \pkg{SIMCOAL2}. It shares its ability to deal with highly complex demographic scenarios, but its simulation approach is radically different. Its focus on performance contrasts with the more relaxed stance of its predecessor. For one thing, it completely abandons \pkg{SIMCOAL2}'s original time-discretized approach in favor of a classical continuous-time approximation. In fact, \pkg{fastsimcoal}'s algorithm is an implementation of SMC', customized to support migration between demes.

Serial sampling capabilities, previously enabled by an extension, are now integrated into the main program. Another addition to \pkg{fastsimcoal} is the possibility of specifying the distributions of various genetic parameters instead of the parameters themselves, facilitating the application of statistical methodologies such as approximate Bayesian computation. Support for recombination hotspots completes the list of new functionalities.

Version 2.0 \cite{Excoffier2021} brings many new features, most of which relate to parameter estimation and inference. Regarding performance, the main improvement involves concurrent computing, with multithreading available for independent simulations. As the implementation is done via \pkg{OpenMP}, this is limited to shared-memory machines. Approximations to transcendental functions, as well as some other unspecified optimizations, also contribute to reducing total computing time.

\pkg{fastsimcoal} is largely backward compatible with \pkg{SIMCOAL2}. They share a similar format for input files. In particular, their input files share the same format, with the exception that \pkg{fastsimcoal} requires specification of polymorphic sites exclusively and the addition of a syntax for haplotype age specification, with omission interpreted as present time.

Simulation studies reported in the original applications note suggest that \pkg{fastsimcoal} produces data similar to \pkg{ms} and \pkg{MaCS} while being consistently faster. Documentation is extensive, with many screenshots and a wide selection of examples. Binaries for major operating systems are distributed directly on the author's website. Unlike \pkg{SIMCOAL2} and most software oriented towards scientific research, source code is not publicly available.

\subsection{\pkg{SHRUB} Family}
This family comprises three related ARG inference pieces of software that exploit heuristics to construct parsimonious genealogies. Their algorithms share a similar aggressive cleaning/reduction step and a simplified recombination procedure. \pkg{SHRUB} and \pkg{beagle} primarily focus on computing bounds on the number of recombination events. However, ancestry topologies are simulated in the process and can be recovered afterward.

\subsubsection{\pkg{SHRUB} \cite{Song2005a}}
\classification{\xg}{\cg}{\cg}{\xg}{\qm}{\qm}

\pkg{SHRUB} is an acronym for ``simulated history recombination upper bound''. It is a heuristic algorithm revolving around a quantity called \emph{recombination weight}. A haplotype $r$ has a recombination weight $w(r \given A - r)$ with respect to a set of haplotypes $A$ if the smallest number of sequences in $A \setminus r$ that $r$ is a mosaic of is $w(r \given A - r) + 1$. From a biological standpoint, this quantity can be interpreted as follows: a haplotype with weight $w(r \given A - r)$ can be coalesced into $A$ after $w(r \given A - r)$ recombinations and $w(r \given A - r)$ coalescence events.

\pkg{SHRUB}'s primary function is to compute an upper bound on the minimum number of recombination events for a sample of haplotypes, which is equal to the minimum of the sums of the recombination weights associated with the ancestries it produces. This sum can be interpreted as the number of recombination events in the best genealogy a naïve parsimonious algorithm can generate. \pkg{SHRUB} comes in two flavors. In its full version, the exact upper bound is computed by branch and bound. Data is represented, at least conceptually, as a matrix, and biological events are equated to row and column deletions. Coalescence and recombination events correspond to the removal of a row, while mutation events correspond to column deletions. The algorithm alternates between two stages; the first greedily performs every currently possible coalescence and mutation events. A row can be removed if it has a recombination weight of zero, meaning that it is equal to some other row in the matrix. This corresponds to a coalescence event. For a column to be a candidate for deletion, it must have no more than one 0 or one 1, a requirement in line with the ISM. The second step involves removing a row and adding its weight to an accumulator before returning to the first stage. When only a single row remains, the accumulator contains the accumulated recombination weight of the ancestry. This quantity is stored in memory, and the algorithm restarts to explore a new ancestry.

The algorithm, as described, requires an exhaustive search to compute the upper bound. As the authors noted, minor branch and bound type modifications significantly increase efficiency. If, at any point, the accumulated weight is greater than the current minimum, the ancestry can be discarded. Similarly, if the current state has been reached in another run of the algorithm, the partial accumulated weights should be compared, and the current run interrupted if its weight is greater. This approach drastically improves performance, reducing the time complexity from $\mathcal O(n!)$ to $\mathcal O(2^n)$. If this speedup is not sufficient, one can use the fast version of the algorithm, where the removed row is chosen to be a minimizer of $w(r \given A - r) + L(A - r)$, the right term being a (possibly approximate) lower bound on the minimal recombination number of $A \setminus r$.

Unfortunately, we were not able to find \pkg{SHRUB}'s \st{binaries nor its} source code. Since the article makes no mention of the language used to program it, we have no way of determining it, although \proglang{C} would be a natural choice. \hl{Binaries are available on the author's website (the link provided in the original paper is no longer valid)}. Similarly, unlike \pkg{HapBound}, the other programs presented in the article, no details about \pkg{SHRUB}'s usage are given, leaving its interface undetermined for the reader. One could reasonably suppose, however, that it is similarly controlled via a command-line interface.

\subsubsection{\pkg{beagle} \cite{Lyngsoe2005}}
\classification{\xg}{\cg}{\cg}{\xg}{\proglang{C}}{CLI}

\pkg{beagle} implement a heuristic designed to find parsimonious ancestries via a branch and bound algorithm. The constructed ancestries are truly parsimonious; they effectively attain the recombination lower bound. \pkg{beagle} assumes the infinite-site model, biallelic markers, and polarized data-that is, the wild allele is known for each marker. Missing data is supported and treated as non-ancestral material.

The algorithm starts by computing a lower bound for the number of recombination events. This is done via a custom algorithm generalizing \pkg{RecMin.c} \cite{Myers2003}. The next step is to attempt to find an ancestry realizing this lower bound, a problem \pkg{beagle}'s algorithm proceeds to solve in a somewhat greedy fashion. First, a reduction step, during which every possible coalescence and mutation event is performed, is repeated until exhaustion. Only then are recombination events considered. A maximum of two consecutive recombination events, which are explicitly restricted to types 1 and 2, are selected, after which another reduction step is executed. This back-and-forth between reduction and recombination is repeated until either a complete ancestry is found, in which case the algorithm finishes, or the number of recombination events exceeds the lower bound. In that event, the bound is increased by one, and the algorithm backtracks. Importantly, \pkg{beagle} computes a new lower bound every time a new state is reached and backtracks early if the minimum number of recombination events required for completion is higher than currently allowed. To avoid re-exploring unfruitful paths, each state along with the associated lower bound is stored in a hash table.

Biological events are formulated as operations on the set of sampled haplotypes. A coalescence event corresponds to the removal of a sequence from that set, while a mutation event is realized by removing a given marker from every haplotype. Recombination events are slightly more complex, as they demand the removal of one haplotype followed by the addition of two sequences. These operations are, of course, subject to restrictions to enforce their biological soundness under the assumptions made by the algorithm. Sequence removal takes place if there is at least one other sequence identical to the one being removed in the set. A marker cannot be removed unless there is at most a single sequence carrying the derived allele. The new sequences resulting from a recombination event must be equal to the old one to the left (respectively right) of the breakpoint and be entirely non-ancestral to its right (respectively left).

\pkg{beagle} is more efficient than \cite{Song2003} since its branch and bound approach avoids enumeration of marginal trees, a claim supported by the simulation studies reported by the authors. Unfortunately, we were unable to find the program's source code or binaries online (the link given in the article is no longer working).

\pkg{beagle} should not be confused with the popular genotype phasing and imputation software \pkg{BEAGLE} \cite{Browning2018,Browning2021}.

\subsubsection{\pkg{KwARG} \cite{Ignatieva2021}}
\classification{\xg}{\cg}{\cg}{\xg}{\proglang{C}}{CLI}

\pkg{KwARG} is described by its developers as parsimony-based, greedy, and heuristic. At first glance, its most striking feature is perhaps its somewhat relaxed adherence to the ISM framework. Technically, it is perfectly capable of producing consistent ancestries in the usual sense. However, unlike the majority of ARG inference software, this is offered as an option; given an appropriate set of parameters, a recombination-free ancestry will be produced. \pkg{KwARG} offers users the possibility to interpolate, so to speak, these two radically different approaches.

Under the hood, \pkg{KwARG}'s algorithm relies on matrix-like data structures, interpreting biological events as operations on the rows and columns of matrices. Sampled haplotypes are stored in a ternary matrix (0, 1, and *) in order to account for non-ancestral or missing material, which are dealt with in the same way. Each iteration of the algorithm begins with repeated application of a cleaning step taken from \cite{Song2003}, which is nothing more than the reduction step of \pkg{beagle} with an additional rule allowing removal of all zero columns. Subsequently, a set of candidate next states is created by generating either a recombination or a (recurrent) mutation event. Restrictions on admissible events are the same as for \pkg{beagle}. After cleaning, a score is attributed to each candidate state in a fashion resembling the A* algorithm \cite{Hart1968}. The cost associated with a given state is a function of user-defined per-event costs and a local estimate of the number of events to termination. The next state is chosen to minimize the cost, but a probabilistic component is introduced in an attempt to compromise between exploration and exploitation.

\pkg{KwARG} compares favorably to exact solutions when computing parsimonious ancestries, both under models that allow for and prohibit recurrent mutations. Remarkably, \pkg{KwARG} can be used to study organisms with a high mutation rate, such as viruses, an avenue explored by the author in a subsequent paper \cite{Ignatieva2022}.

Source code, binaries for major operating systems, and small utility programs are distributed via GitHub.

\subsection{\pkg{Margarita} Family}
This category includes programs aiming at inferring so-called ``plausible'' or ``minimal'' ARGs via a heuristic approach related to that implemented by \pkg{Margarita}. \pkg{Margarita} itself is in some respects similar to \pkg{SHRUB}. According to the authors of the former software, these resemblances are coincidental. Consequently, we have decided to split them into distinct families.

\subsubsection{\pkg{Margarita} \cite{Minichiello2006}}
\classification{\xg}{\xs}{\xs}{\xg}{\proglang{Java}}{CLI}

The primary motivation behind \pkg{Margarita} is the fine mapping of diseases via inferred ARGs. Even though the implemented algorithm is heuristic-based, the authors clearly state that their guiding principle is not the minimization of the number of recombination events. They aim for constructed ARGs to be ``plausible'', a term that they unfortunately do not define further. In any case, their method is efficient, allowing for sample sizes in the thousands and sequences of hundreds of SNPs. The authors suggest that \pkg{Margarita} could be used on the genome scale, albeit in a windowed fashion.

The heuristic is guided by a few principles. Coalescence events are restricted to type A. Regarding mutations, back or recurrent ones are prohibited. In consequence, mutation events for a given marker can only occur when there is a single sequence having the derived allele, and must occur on that sequence's parent. Recombination events are chosen to allow immediate coalescence, which is indeed performed directly afterwards. It may be necessary to actually perform two recombination events before a coalescence is possible, a possibility \pkg{Margarita} is designed to handle. In any case, recombining sequences are chosen to maximize shared material. Under these rules, coalescence and mutation events are produced greedily until exhaustion. At that point, a recombination event is sampled. This process is iterated until the sample's MRCA is reached.

From a practical standpoint, \pkg{Margarita} can deal with missing and unphased data. Absent SNPs are imputed \textit{ad hoc} by basically being treated as non-ancestral material. The same idea, with additional restrictions, is used for unphased markers.

\subsubsection{\pkg{ARG4WG} \cite{Nguyen2017}}\label{arg4wg}
\classification{\xg}{\xs}{\xs}{\xg}{\proglang{C++}}{CLI}

The main purpose of \pkg{ARG4WG} is to enable analysis of genome-sized datasets. Its authors acknowledge the performance of \pkg{Margarita} while noting that it is still not fast enough to handle ``thousands of whole chromosome samples''. Their software is a reimplementation of \pkg{Margarita}, incorporating a modification to the recombination step.

The recombination step of \pkg{Margarita}'s algorithm selects an haplotype to maximize the amount of shared ancestral material in the subsequent coalescence events. Since the shared segment can be flanked by unshared ancestral material, up to two recombination events may be necessary before coalescence is possible. \pkg{ARG4WG}'s algorithm is similar but restricts recombination events even further: every single event must \emph{immediately} result in a coalescable sequence. This is enforced by selecting candidates on which the shared material is positioned at one of its ends or at least only separated from the end by a non-ancestral segment. \pkg{ARG4WG}'s algorithm is justified by a lemma stating that given a set of reference haplotypes and an additional haplotype to be coalesced into it, a greedy left-to-right procedure repeatedly generating recombination events and coalescing the resulting sequence with ancestral material to the left of the breakpoint in a way that maximizes the amount of shared ancestral material minimizes the number of recombination events. This result is similar to the lemma 4.2 of \cite{Song2005a}. In addition, the search for haplotype pairs with shared ends can be conducted in parallel, which further speeds up execution.

Although biological realism is not a priority, \cite{Nguyen2017} shows results in line with \pkg{Margarita}, and the results of their application to the Gambia dataset agree with other analyses. Regarding performance, the results are undoubtedly impressive. For instance, \pkg{ARG4WG} was able to construct an ARG for a sample of 5560 haplotypes containing data for the whole chromosome 11 in just over three hours using 16 threads.

\pkg{ARG4WG}'s source code is distributed on GitHub. A short readme containing compilation and execution instructions is available.

\subsubsection{\pkg{GAMARG} \cite{Thao2019}}\label{gamarg}
\classification{\xg}{\xs}{\xs}{\xg}{\proglang{C++}}{CLI}
\pkg{GAMARG} aims to reduce the number of recombination events even further than its predecessor \pkg{ARG4WG}, to which it is closely related. The main modification is the integration of the well-known \emph{four gametes test} \cite{Hudson1985} into its algorithm. Since this test operates pairwise on markers, it does not scale to the sample sizes \pkg{GAMARG} is intended to handle. Consequently, it actually implements a windowed and weakened version of the complete test.

This addition to \pkg{ARG4WG}'s algorithm has important consequences. One of the important features of \pkg{ARG4WG} is that the number of live sequences never increases. Any recombination event added to the ancestry is chosen to enable a specific coalescence event that follows immediately. This is no longer the case for \pkg{GAMARG}. After every coalescence and mutation events have been exhausted, a ``classic'' recombination event, that is one not followed by a coalescence, is generated if there is at least one sequence bearing a pair of incompatible sites.

Simulation studies and application to real datasets suggest a slight reduction in the number of recombination events. However, execution time is considerably above that of \pkg{ARG4WG}.

\subsubsection{Other Programs}
\begin{itemize}
\item REARG \cite{Thao2017}: Consider both the ``similarity between sequences'' and the ``length of sequences'' when generating recombination events. Since we could not access the paper or the software's source code, our summary is based on \cite{Thao2019}.
\end{itemize}

\subsection{\pkg{MaCS} Family}
This section groups simulators implementing a flexible ARG construction algorithm related to that of \pkg{MaCS}.

\subsubsection{\pkg{MaCS} \cite{Chen2008}}\label{sec:macs}
\classification{\cg}{\cg}{\xs}{\cg}{\proglang{C++}}{CLI}

The main appeal of exact CWR simulators such as those belonging to the \pkg{ms} family is the possibility of obtaining accurate samples from the distribution of interest. In addition, some of them even allow for rich demographic and even selective models. However, before \pkg{msprime}, they were more or less relegated to the rank of academic tools due to their difficulty in keeping up with the rapidly growing scale of real-world data. SMC approximations are definitely less limited in that regard, but trading biological realism backed by mathematical rigour for computational performance gains can be a tough sell. \pkg{MaCS} intends to be a middle ground between the two approaches by implementing a variable approximation of the CWR. The main feature of the SMC algorithm is its Markovian nature: any marginal tree is independent of any previous one given one in between. This allows for efficient simulation algorithms since only the current tree needs to be considered when simulating the next one. From the point of view of sequential approximations, \pkg{MaCS}' algorithm can be summarized as an SMC-type simulator that generalizes the concept of marginal \emph{tree} to that of marginal \emph{graph}. While the SMC generates the next tree conditional on the current position, \pkg{MaCS} has the ability to do so conditional on the history of an \emph{interval} of positions. The right endpoint of this interval is the current position and its length is controlled by the user. Graphically, this interval's ancestry is described by the subgraph induced by the subset of vertices having ancestral material within it. In general, for an interval, there is no guarantee that this subgraph is a tree, hence our expression ``marginal graph.'' The number of marginal trees $k$ included in any given marginal graph is determined by a sequence length $h$ specified in the program's invocation. $k$ is chosen to represent the ancestry of an interval of position with width $h$. Since the number of recombination events within a specified interval is random, this choice is approximate.

We can infer from the fact that recombining sequences are allowed to recoalesce on the current marginal graph without further restriction that both types of coalescence events are possible providing a sufficiently large $h$. Recombination events' locations are constrained to the marginal tree associated with their position. In consequence, type 2 events are impossible. By the same reasoning, non-informative events are not simulated either.

Regarding code, data structures constitute an interesting departure from \pkg{ms}, which encodes ARGs as sequences of trees. \pkg{MaCS}' approach is to store the complete ARG as a graph. The CLI is mostly similar to \pkg{ms}' and an auxiliary program called \pkg{msformatter} that turns \pkg{MaCS} output into one compatible with Hudson's \pkg{ms} is provided.

\subsubsection{\pkg{SC} \cite{Wang2014}}
\classification{\cg}{\cg}{\cg}{\cg}{\proglang{C++}}{CLI}

The only feature missing from \pkg{MaCS} in order to fully interpolate between Hudson's and SMC algorithms is the capability to generate type 2 recombination events. This is what \pkg{SC} brings to the table. Based on version 0.4e of \pkg{MaCS}, both software implement largely similar algorithms, with \pkg{SC} making one important addition to the recombination step. In line with other sequential simulators, \pkg{MaCS} select recombination locations uniformly on some subgraph of the ARG conditional on their position on sequences. While this is also true of \pkg{SC}, this is not the whole story. After one such event has been produced, two things can happen depending on the location of the subsequent recoalescence event. If it is located on the current marginal tree, the algorithm iterates just like \pkg{MaCS} would. If, on the other hand, recoalescence occurs on an older marginal tree, a chain of successive recombination and recoalescence events is started. The next recombination is simulated from the previous recoalescence's location following a set of rules. This process continues until coalescence with the current marginal tree occurs.

\subsubsection{\pkg{SC-sample} \cite{Wang2014}}\label{sc-sample}
\classification{\cg}{\cg}{\cg}{\xg}{\proglang{C++}}{CLI}

Modified version of \pkg{SC} integrating a minimum mutation number (MMN) algorithm as well as various constraints that allow ARG inference from a sample of haplotypes. Their MMN algorithm is an edge labelling scheme allowing efficient computation of the minimum number of mutation events that would have been necessary for a marginal tree to have given rise to a sample's specific marker. If this number is larger than one, then the marginal tree cannot represent this marker's history assuming the ISM. Otherwise, it is said to be \emph{consistent} with the marker. A high-level description of the algorithm is as follows: start by constructing a consistent marginal tree for the leftmost marker, which is easily done. Apply the MMN algorithm sequentially to each following marker. When a marker $m$ is found to be inconsistent with the last marginal tree, generate recombination events between that marker and the last one (or the beginning of the sequence) until a new marginal tree consistent with $m$ is built. Iterate until every marker is associated with a consistent tree.

While the description of the algorithm in \cite{Wang2014} is slightly unclear with respect to the recombination positions' distribution, a glance at the code confirms that each one is uniformly distributed between the last one (or the position of marker $m - 1$) and the $m$\textsuperscript{th} marker's position. This choice of a support-bounded distribution is a departure from standard theory necessary when inferring ARGs. It would be naive to hope for a sample to be produced in reasonable time otherwise. The distributions of recombination and recoalescence locations are similarly bounded versions of their classical counterparts. Recombination edges are constrained to be mutation edges and to recoalesce in a way that decreases the number of mutations.

\subsection{Tree Scan Family}
ARG inference based on the so-called tree scan \cite{Song2005} method.

\subsubsection{\pkg{RENT} \cite{Wu2011}}
\classification{\xg}{\cg}{\cg}{\xg}{\proglang{C++/Java}}{CLI}

\pkg{RENT} aims to improve on earlier methods. The ``tree scan'' method
\cite{Song2005}, which is the focus of the first part of the paper, can produce
ancestries with a number of recombination events attaining its lower bound. This
method works well, but it does not scale up. It is essentially an exhaustive
search, making it practically impossible to deal with sample sizes of more
than nine haplotypes. \pkg{RecPars} \cite{Hein1993} attempts to overcome this
limitation using heuristics. Some of its drawbacks include its assumption that
marginal trees are rooted and binary. While this is a fairly common assumption,
the author notes that such methods heavily depend on the unavoidable choice
of an initial tree, which is generally arbitrary. It also limits the number of
recombination events between two markers to one and is unable to infer more
than one ancestry.

Central to \pkg{RENT}'s algorithm is the phylogenetic concept of a split, which
in that context is nothing more than a bipartition of a set of haplotypes. If
each haplotype is associated with the leaf of a tree, a split is simply an edge;
removing any edge of a tree results in a graph with two components. A split
with a singleton component is deemed trivial and is non-informative because
it appears in any tree. In general, the smaller the set of trees for which a
given split appears, the more informative the split. For instance, a star graph
is the least informative topology since every split is trivial. By contrast, a
completely unbalanced binary tree is maximally informative since every partition
is possible. \pkg{RENT} starts with a set of ``trivial'' marginal trees, one
for each marker. A trivial tree has a single non-trivial split partitioning
haplotypes based on their status with respect to a marker. Those trees are
iteratively refined towards topologies with more and more non-trivial splits.
This is done through heuristic rules that take into account not only the tree
itself, but its neighbors as well.

A few years later, the author published an extension called \pkg{RENT+}
\cite{Mirzaei2016}, which builds on the same split refinement ideas. The main
improvement concerns the initial set of marginal trees, which is constructed
based on the distance between sampled haplotypes. The metric used is nothing
more than a windowed and normalized version of Hamming's distance. In the
original program, singleton markers - that is, markers for which only one
haplotype bearing the derived allele exists in a sample - were simply ignored;
any information they may contain about inter-sequence similarities is not
considered.

\pkg{RENT} is written in \proglang{C++}. We were unable to locate it online.
\pkg{RENT+} is written in \proglang{Java} and is available on GitHub.

\subsubsection{\pkg{ARBORES} \cite{Heine2018}}\label{arbores}
\classification{\cg}{\xs}{\xs}{\xg}{\proglang{C}}{CLI}

The main idea of \pkg{ARBORES} is to scale up the tree scan procedure by
applying it to overlapping subsegments of haplotypes. We classify it as
model-based regardless, as the heuristic serves as the foundation for a Metropolis-Hastings
MCMC sampler designed to generate draws from the SMC distribution. Densities are
derived from a hidden Markov model (HMM) with marginal trees and haplotypes as
hidden and observed states respectively.

Proposals are obtained by inferring ancestries for each subsegment conditional
on the marginal trees of its leftmost and rightmost markers, a procedure known
as ``bridging''. Importantly, subsegments can be processed concurrently. It
begins with the computation of candidate topologies using a tree scan-inspired
procedure \cite{Song2005}, which involves the iterative modification of a
marginal tree by subtree-prune-and-regraft (SPR) operations. As the name
suggest, an SPR operation consists in cutting a branch of the ancestry and
reattaching it elsewhere. From a biological standpoint, it is equivalent
to a pair of consecutive recombination and recoalescence events. As in
\cite{Hein1993}, a parsimony constraint limiting the number of SPR operations
between adjacent sites to at most one is enforced. The bridge is completed by
drawing latitudes for free vertices, that is, those absent from either terminal
tree.

In theory, the search involved in the bridging procedure should be exhaustive.
As it turns out, the computational burden is prohibitive, and actual
implementation relies on a two-pass approximation. First, a bridge with the number of SPR operations between non-segregating and segregating
sites restricted to a maximum of 0 and 2, respectively, is selected at random. Trajectories for
which fulfillment of this constraint is not possible are simply discarded.
To introduce variability, the set of candidates is augmented by subjecting
topologies obtained in the first pass to additional SPR operations. Each is
processed from left to right by performing an SPR operation at sites for which
none were necessary in the first pass.

In addition to its inference algorithm, \pkg{ARBORES} also implements a
tailor-made branch-and-bound consistency-checking procedure. It is provably
exact, assuming no more than one SPR operation between segregating sites.
Unfortunately, this proof does not hold for greater numbers of operations.

\pkg{ARBORES}'s approach is certainly noteworthy, particularly due to its
divide-and-conquer approach to model-based ARG inference, which enables
concurrency while supporting a rich dependence structure. With that bieng said,
the authors' own numerical experiments indicate that the current implementation
of the algorithm does not scale.

\pkg{ARBORES} accepts input in a custom format documented in a PDF available in
the GitHub repository from which it is distributed. The documentation contains
concise information about other aspects of the software.

\subsection{\pkg{ARGWeaver} Family}
This family is composed of programs implementing threading-based algorithms.

\subsubsection{\pkg{ARGweaver} \cite{Rasmussen2014}}\label{argweaver}
\classification{\cg}{\xs}{\xs}{\xg}{\proglang{C++}}{CLI}

\pkg{ARGweaver}'s approach to ARG inference is radically different from its predecessors'. The objective of inferring ancestries for a set of haplotypes is reframed as solving a sequence of \emph{threading problems}. In essence, it involves constructing an ARG for $n$ haplotypes conditional on an ARG inferred for $n - 1$ other haplotypes under Markovian assumptions regarding the recombination process. Due to the notoriously large size of the space of possible ARGs, this is easier said than done. \pkg{ARGweaver} addresses this difficulty by implementing a tailor-made algorithm revolving around a space-time discretized SMC-type model called the \emph{discretized sequentially Markov coalescent} (DSMC). Dependency between marginal trees is modeled as in the original SMC, but the locations of coalescence and recombination events, as well as the positions of the latter, are constrained to a discrete grid. The DSMC can be viewed as an approximation to the SMC and is designed in such a way as to approach it (under some mild assumptions) as the grid is refined. The main drawback of this model is that it allows for zero-length branches with all related rounding concerns. These are addressed heuristically by \pkg{ARGweaver}. In any case, these assumptions allow for a tractable expression of the joint density of the ARG and haplotypes as the complete data likelihood function of a HMM.

A threading problem is solved by adding recombination and recoalescence events to the current graph. Concretely, this amounts to adding one branch between the sequence to be threaded and each marginal tree. The sequence of locations for these new branches on the current graph is called the \emph{coalescence threading} of the new sequence. A \emph{recombination threading} describing related variations between marginal trees is defined similarly. One important optimization is a consequence of the heavy restriction imposed on the recombination threading by the coalescence threading: it is considerably easier to sample the former conditional on the latter than to sample them jointly. This is the approach implemented in \pkg{ARGweaver}. Moreover, this procedure can be carried out in parallel because recombination events, within this framework, are conditionally independent.

Entire ARGs can be inferred through threading. After a first trivial graph containing a single sequence is constructed, the remaining sequences are threaded one at a time until completion. This is perhaps the most intuitive use case for \pkg{ARGweaver} and can be done efficiently. The main drawback of this approach is that the graph built when threading sequence $k + 1$ is sampled conditional on the first $k$ sequences rather than on the whole sample, making the distribution of the resulting ARG invalid. The authors' recommended solution to this problem is to use sequence threading as the starting point for a Markov chain Monte Carlo (MCMC) sampler. The basic approach is to build an initial ARG using the method described previously and subsequently remove and rethread sequences. Again, this can be done efficiently. However, the impact of rethreading a single sequence on the graph is generally fairly local, leading to poor mixing. To improve the sampler's quality, \pkg{ARGweaver} implements subtree sampling. Sampling a whole subtree is not an easy task, mainly because subtrees are not constant along sequences. Consequently, \pkg{ARGweaver}'s subtree sampler relies on a fairly involved subtree selection algorithm, which takes a toll on performance.

In short, reframing ARG inference as a sequence of threading problems has the potential for significant performance gains over traditional methods, since such problems can be solved efficiently. Nevertheless, obtaining a satisfactory sample-that is, one whose distribution is ``close enough'' to that of the CWR-from repeated rethreading is not straightforward.

\pkg{ARGweaver} is distributed with a handful of utilities in the form of individual programs. The sampler itself is a standalone piece of software called \pkg{arg-sample} which takes optionally compressed \code{fasta}-formatted files as input. A more compact custom file format called \code{sites} can also be used. Simulations will be started from scratch by default, but it is possible to specify an ARG as an initial value or even resume execution from a partially built graph. Simulation results are stored as Newick-encoded marginal trees along with data about recombination events. TMRCA and recombination rate can be estimated from generated ancestries using \code{arg-extract-tmrca} and \code{arg-extract-recomb}. Facilities for ARG visualization are also provided. Conveniently, \pkg{ARGweaver}'s distribution also includes \pkg{arg-sim} which can be used to generate samples of sequences under both SMC and classical CWR models along with DSMC. Implementations are custom.

\pkg{ARGweaver} has been subsequently modified to include demographic models, an extension known as \pkg{ARGweaver-D} \cite{Hubisz2020}.

\subsubsection{\pkg{ARG-Needle} \cite{Zhang2023}}\label{arg-needle}
\classification{\xg}{\xs}{\xs}{\xg}{\proglang{C++}}{CLI*}

\pkg{ARG-Needle} addresses \pkg{ARGweaver}'s limitations with a novel model for the TMRCA distribution, along with a marker-based distance heuristic. The algorithm is divided into three steps, the third of which involves threading a new sequence into the ARG. These steps are repeated until the complete ARG is built, that is, until every sequence in the sample has been threaded. The heuristic is applied in the first step, identifying the most similar haplotypes among those already threaded. To account for recombination, the search is localized to subdivisions of sequences, the length of which is a user-defined parameter. This results in a list of the most similar haplotypes for each subdivision. A hashing scheme is used to speed up this process. In the second step, the TMRCA between the sequence to be threaded $s$ and each sequence computed in the previous step is estimated. Again, this estimation is performed per subdivision, using the Ascertained Sequentially Markovian Coalescent algorithm \cite{Palamara2018}, which revolves around an HMM model of the haplotypes where the hidden state is made up of TMRCA intervals. Coalescent threading for $s$ is computed in the third step using the minimum estimates from the previous step. Unlike \pkg{ARGweaver}, recombination threadings are not computed. Also, \pkg{ARG-Needle}'s implementation of the threading procedure is ``ARG-wide'' (e.g., it does not consider each marginal tree individually), which increases its efficiency. A post-processing step called \emph{ARG normalization}, \hl{which amounts to quantile renormalization on the heights of internal nodes based on trees sampled from the demographic prior, may} allow\st{s} for a reduction of bias in the TMRCA estimates.

Despite both pieces of software being threading-oriented, \pkg{ARG-Needle} and \pkg{ARGweaver}'s approaches to ARG inference are radically different. The former prioritizes performance, heavily relying on heuristics and largely leaving \pkg{ARGweaver}'s distribution validity concerns unaddressed\st{. While this is by no means a bad thing, one should be cautious when presenting} \pkg{\st{ARG-Needle}} \st{as a more efficient ``version'' of} \pkg{\st{ARGweaver}}. That being said, recent work \cite{Bisschop2025} suggests that satisfactory likelihoods can be formulated for ARGs that do not include recombination nodes under the SMC. However, the fundamental issue of the threading approach remains: the $k$\textsuperscript{th} threading is independent of the remaining $n - k$ unthreaded haplotypes. The existence of a likelihood function is unrelated to the statstical properties of the inferred ancestry.

In addition to the CLI interface, a \proglang{Python} API to the underlying library (called \pkg{ARG-Needle-lib}) is also available and distributed via PyPI.

\subsubsection{\pkg{Threads} \cite{Gunnarsson2024}}\label{threads}
\classification{\cg}{\xs}{\xs}{\xg}{\proglang{C++}}{CLI}

\pkg{Threads} is closely related to \pkg{\st{ARGneedle} \hl{ARG-Needle}}. Both were designed in Pier Palamara's lab within a short time interval, and the papers presenting each software have many authors in common. Both programs are designed to solve threading problems using heuristics and are very fast at doing so, enabling genome-scale ARG inference. However, a major difference between the programs lies in \pkg{Threads}'s use of the Li-Stephens model (LSM) \cite{Li2003}. The popular framework, which is mentioned by \pkg{ARGweaver}'s authors, is related to threading, as it revolves around a factorization of the haplotypes' joint density of the form
\begin{equation*}
  f(h_1, \ldots, h_n \given \rho) =
  \prod_{k = 1}^n f(h_k \given h_1, \ldots, h_{k - 1}, \rho)
\end{equation*}
where the $h_i$s are haplotypes and $\rho$ is a recombination parameter. \pkg{Threads} extends this expression slightly to include a mutation parameter, which we denote by $\mu$ in this section. In any case, the fundamental idea is to view the $k$\textsuperscript{th} haplotype as a \emph{mosaic} of the first $k - 1$ sequences which constitute a \emph{reference panel}. Multiple problems, such as finding recombination breakpoints, can be solved using standard approaches for hidden Markov models within that framework.

The \st{PAC} \hl{LSM} model itself is sufficient to compute a coalescent threading, a fact noted by the authors. However, the large search space of possible threadings is problematic from an efficiency standpoint. For this reason, the set of candidates for new coalescence events is first reduced in a pre-processing step via a tailor-made algorithm based on the positional Burrows-Wheeler transform \cite{Durbin2014}. Thanks to this heuristic, only candidates similar to the new haplotype, with respect to the Hamming distance, are considered. The implementation is efficient as it revolves around operations on red-black trees. \hl{Indeed, numerical experiments presented by the authors show that} \pkg{\hl{Threads}} \hl{is generally faster than} \pkg{\hl{ARG-Needle}}\hl{, often by an order of magnitude.}

A new haplotype $h_{n + 1}$ is threaded into the graph according to the sequence of events maximizing $f(h_{n + 1} \given h_1, \ldots, h_n, \rho, \mu)$. $h_{n + 1}$ is associated with a word in $\{ 1, \ldots, n \}^m$ representing the pairwise similarity between two haplotypes for each marker. The number $k$ at position $p$ indicates that the new haplotype is most similar to sequence $k$ at the $p$\textsuperscript{th} marker, given a pair of recombination and mutation rates. Of course, runs of similar letters are expected; these are called \emph{Viterbi segments} and are conceptually related to the widespread notion of \emph{IBD segment}, although they are not interchangeable. Each Viterbi segment making up a haplotype constitutes a threading unit, acting as a proxy for recombination events (see section 5 of supplementary notes). In any case, maximizing the conditional density of a new haplotype is computationally challenging, as the classical algorithm involves $\mathcal O(mn^2)$ evaluations. \pkg{Threads} significantly reduces this number. First, the pre-processing step described above is used to reduce the number of candidates to $l$, which is chosen to be small, possibly orders of magnitude smaller than $m$. While this is helpful, the heart of \pkg{Threads}'s algorithm is a branch and bound approach based on certain characteristics of the recombination process, lowering the complexity of the so-called Threads-Viterbi step to linear in $n$. Since there are $m$ haplotypes to be threaded, the complexity of this step is $\mathcal O(mln)$. A final source of speedup comes from the fact that each of the $m$ optimization problems is independent of the others, making computation perfectly parallelizable.

By default, \pkg{Threads} ancestries cannot be considered to be inferred from a sample in the usual sense. An optional post-processing step, which estimates the age of each mutation and alters the ARG topology to render it sample-consistent, is available to the user via a simple command-line switch. \hl{Optionally, mutation ages can be specified by the user.}

\pkg{Threads} takes three files as input. The first two files should provide genotype data and genetic maps in the form of \pkg{Plink2} and \pkg{SHAPEIT} formatted data, respectively. The third file should supply information about demography in a custom two-column format. The first column denotes a number of generations in the past, while the second column indicates the corresponding effective population size. The output is in a custom format (.threads file) but can be easily converted to ARG-Needle format using a built-in utility. It is worth mentioning that these files can later be used for genotype imputation by \pkg{Threads} itself. \hl{This is an important use case and another distinctive feature with respect to ARG-Needle}

\pkg{Threads} is coded in \proglang{C++} and \proglang{Python}. It is made available through the PyPI index. The ARG construction process is controlled via a well-documented command-line interface. Despite a significant portion of the software being written in \proglang{Python}, there is currently no publicly documented \proglang{Python} API.

\subsubsection{\pkg{SINGER} \cite{Deng2025}}\label{singer}\label{singer}
\classification{\cg}{\xs}{\xs}{\xg}{\proglang{C++}}{CLI}

Similar to \pkg{ARG-Needle}, \pkg{SINGER} main goal is to improve upon \pkg{ARGweaver} by reducing inference time, although it does not, for the most part, resort to heuristics for that purpose. As stated above, the main performance bottleneck in \pkg{ARGweaver} comes from its MCMC sampling step, which, while not mandatory, is necessary for full posterior sampling. It is therefore not surprising that one of the main innovations of \pkg{SINGER} is a simplification of the MCMC sampling step. The major source of complexity of \pkg{ARGweaver}'s subtree sampling algorithm comes from the requirement for cuts to spatially span the totality of the ARG's genome positions. While mixing is improved, the procedure's sophistication is high, and its computational burden heavy. \pkg{SINGER}'s subgraph pruning and regrafting operation represents a tradeoff, allowing efficient cuts on internal branches at the cost of reduced spatial span. At each step, the position of the next cut is sampled to the right of the interval covered by the last one. Its location on the associated marginal tree is sampled uniformly. This position is then stretched into an interval in both directions. Stretching in a given direction stops when either the end of the sequences is reached or a position deemed equivalent to the cut is encountered. Pruned subgraphs are regrafted via threading by approximate sampling from the posterior distribution, which leads to high acceptance rates. This means that, in addition to \pkg{SINGER}'s sampling procedure being more efficient than \pkg{ARGweaver}'s subtree sampling, a reduction in mixing times is expected. This is confirmed by simulation studies presented in \cite{Deng2025}, which report a total speedup of around 400x, largely due to improved mixing.

Another source of improvement with respect to \pkg{ARGweaver} is its approach to the threading problem. Even though both software rely on hidden Markov models, \pkg{SINGER} brings once again a simpler solution by considerably reducing the number of hidden states. Instead of considering the complete set of discretized recombination and coalescence points, a two-step procedure is implemented. These steps are known as \emph{branch sampling} and \emph{time sampling}. In the first step, a coalescence branch for the haplotype to be threaded is sampled. This is done efficiently via a deterministic approximation of coalescence times, which is used to compute two quantities. The first is a location on the coalescence branch called the \emph{representative joining time}. It serves as a provisional coalescence location used to compute mutation probabilities. The second is the probability of coalescing with the branch under the coalescent process. It appears, along with the representative joining time, in expressions for recombination probabilities. In HMM parlance, recombination and mutation events correspond to transitions and emissions, respectively. However, this is not the whole story. \pkg{SINGER}'s approach to the threading problem is unique in that single markers do not constitute the unit of analysis (referred to as \emph{locus} in \cite{Deng2025}). For probability computation purposes, they are grouped together in equally sized bins, effectively reducing the number of loci and, consequently, the computational burden. Additionally, the number of recombination events between two loci is restricted to be at most one, giving rise to a second transition type: instead of generating a new recombination event, the new haplotype can join a previous coalescence threading, a process called \emph{recombination hitchhiking}.

\pkg{SINGER}'s restriction on the number of recombination events is very strong. Indeed, in combination with its adherence to the SMC paradigm, it is too strong in the following sense: the set of possible threadings is not absolutely continuous with respect to that of possible ARGs. This is problematic because threadings not associated with a valid ARG could possibly be sampled. To remedy this issue, the authors introduce a distinction between \emph{full} and \emph{partial} branches. A branch is said to be partial if it has been bisected by a recombination or coalescence event; branches that are not partial are full branches. Strictly speaking, it is possible to generate a sequence of marginal trees under the SMC solely by keeping track of full branches. In fact, the previous statement is something of a threading-centric formulation of the SMC. Indeed, recombination and coalescence operations are often described in a somewhat metaphorical language. For instance, representing a coalescence event in an ARG is often described as bisecting the coalescence edge. This is generally helpful, but sometimes obscures important details. From a graph theoretical standpoint, the actual operations involved are addition and deletion of elements in the ARG's vertices and edges sets according to certain rules. It is possible to formulate the set of rules associated with SMC as a subset of that of standard CWR; these restrictions have to do with not adding new vertices or partial branches. In any case, in order to allow recombination hitchhiking, \pkg{SINGER} has to depart from strict SMC and keep track of partial branches. Theoretical complexity improvements granted by approximating the CWR are lost, but can be mostly recovered in practice. By constant pruning of low probability branches, the state space size can be kept close to its theoretical minimum, according to the authors.

Threading computation is followed by a rescaling step. Time is discretized into intervals. For each interval, a scaling factor based on the expected number of mutations is computed. The procedure is similar to \pkg{\st{ARGneedle} \hl{ARG-Needle}}'s normalization step; the key difference being that \pkg{SINGER} does not assume any demographic model.

\pkg{SINGER} is controlled via a CLI interface. It accepts input in the form of standard VCF files. Its output can easily be converted into tskit format using an included \proglang{Python} script. Two scripts are also provided for output analysis. The first computes summary statistics indicative of the MCMC sampler's convergence. One of these, the number of non-uniquely-mapped sites, is reported for simulations in \cite{Deng2025}. The other computes the pairwise coalescence times with respect to a focal haplotype. In addition, two scripts are provided for streamlining concurrent execution of \pkg{SINGER} on regions of input data. The first divides haplotypes into regions of a predetermined length; graphs are then constructed in parallel for each of these. The other allows the user to select a subset of regions for ancestry inference.

\subsection{Others}
\subsubsection{\pkg{SNPsim} \cite{Posada2003}}
\classification{\cg}{\cg}{\xs}{\cg}{\proglang{C}}{CLI}

\pkg{SNPsim}'s primary feature is the ability to simulate recombination hotspots. The user must provide a base recombination rate, the expected number of hotspots, and a "hotspot" recombination rate. The program begins by selecting random locations for hot and cold spots, along with a specific rate for each. \pkg{SNPsim} then performs standard CWR simulation while accounting for the variable recombination rate.

Demographic scenarios with varying population sizes are supported. Complex mutation models are possible. Recurrent mutations can be simulated using a biallelic Jukes-Cantor model \cite{Jukes1969}. Mutation models and rates can be mixed along haplotypes.

Although the article does not mention the types of events supported, the code in the function \code{CalcIndividualG} indicates support for type B recoalescence. Type 2 recombination events are out of scope.

\pkg{SNPsim} is written in \proglang{C} and is publicly available through Google's Code Archive.

\subsubsection{\pkg{FastCoal} \cite{Marjoram2006}}\label{fastcoal}
\classification{\cg}{\xs}{\xs}{\cg}{\proglang{C++}}{\qm}

This is the Marjoram and Wall \proglang{C++} implementation of the SMC and SMC' algorithms. As a reminder, from an algorithmic perspective, SMC's main idea is to delete the recombination branch above the recombination location \emph{before} generating a recoalescence event. SMC' slightly increases its support by deleting the branch \emph{after} the recoalescence event, thus allowing recombination events to be effectively cancelled.

Ancestries produced by \pkg{FastCoal} via SMC' are reportedly similar to those constructed via the original SMC algorithm, but with a distribution more similar to that of \pkg{ms}, used by the author as the reference CWR simulator. This conclusion is based on comparisons of the TMRCAs distributions and the patterns of linkage equilibrium between the two software. As expected, \pkg{FastCoal} is overwhelmingly faster than \pkg{ms} and more memory efficient.

We were unable to find either \pkg{FastCoal} source code or binaries on the internet.

\subsubsection{\pkg{GENOME} \cite{Liang2007}}
\classification{\cg}{\cg}{\cg}{\cg}{\proglang{C++}}{CLI}

Although it proceeds backward in time, \pkg{GENOME} is not, strictly speaking, a CWR simulator since it does not resort to continuous-time approximations. This design decision is motivated by diminishing returns as the length of the haplotypes to be simulated increases. Indeed, simulation studies reported in the paper indicate runtimes inferior to those of \pkg{ms} and \pkg{cosi} for long chromosomes. The implemented scheme is to simulate every generation, working backward in time from the present. In addition to being more time-efficient, discrete generations allow for fine control over demographic events and simultaneous events such as coalescence between more than two sequences. ISM is assumed throughout the paper. Variable recombination rates are supported, enabling the simulation of hotspots. The algorithm is implemented using sparse matrices, which are reused at each generation for efficient memory management. All informative events are supported, as revealed by the code of the \code{SampleParent} function.

Format of the output files is similar to \pkg{ms}. Simulation studies indicate that the output is in line with theory and similar to \pkg{ms}. \pkg{GENOME} is programmed in \proglang{C++} and available online, although not at the address provided in the paper. The interested reader can find its source code along with a manual and precompiled binaries for major operating systems at \url{https://csg.sph.umich.edu/liang/genome/download.html}.

\subsubsection{\pkg{TMARG} \cite{Wu2008}}
\classification{\xg}{\cg}{\cg}{\xg}{\proglang{C++}}{\qm}

\pkg{TMARG} is based on lemma 4.1 of \cite{Wu2008}, an extension of lemma 3.1 of \cite{Song2005a}. An ARG may or may not contain haplotypes absent from its sample; those are known as \emph{Steiner sequences}. An ARG without Steiner sequences is called \emph{self-derived}. A sample for which there exists a self-derived ARG is called \emph{self-derivable}. Lemma 4.1 states that if a sample is self-derivable, every minARG for that sample is self-derived.

\pkg{TMARG} actually implements two different algorithms, both based on parsimony-centric heuristics. The first algorithm presented in the article relies on the fact that sampling uniformly over the space of self-derived ARGs can be done efficiently. Unfortunately, this approach has two issues that greatly reduce its usability in real-world scenarios. First, it requires counting the total number of possible minARGs for a given sample. This can be achieved using a fairly simple procedure, albeit in a time exponential in the size of the sample. An algorithm is described in the paper and implemented in \pkg{TMARG}. Once this is accomplished, however, constructing an ARG is an operation with quadratic complexity.

The second flaw of this approach is perhaps the more obvious: it is only applicable to self-derivable samples. If the sample of interest does not meet this requirement, one strategy is to augment it with Steiner sequences to make it self-derivable. This is not practical in general, as the search space for Steiner sequences grows exponentially with the number of markers. Heuristics aiming at facilitating exploration of this space are presented in the papers, but for larger datasets, \pkg{TMARG} implements an alternative algorithm called \emph{pathway sampling}. The idea is to first draw a permutation of $1, \ldots, n$ where $n$ is the number of sampled haplotypes. Then, a sequence of events generating haplotype $i$ involving only haplotypes 1 up to $i - 1$ is computed for each $i$, and the ARG is built accordingly. Pathway sampling is related to \pkg{SHRUB}'s algorithm.

\pkg{TMARG} is coded in \proglang{C++}. Unfortunately, its source code is not publicly available online.

\subsubsection{\pkg{SLiM} \cite{Messer2013}}
\classification{\cg}{\cg}{\cg}{\cg}{\proglang{C/C++}}{DSL*}
\pkg{SLiM}, which stands for ``Selection on Linked Mutations,'' is an agent-based forward-in-time simulator based on the Wright-Fisher model. Its stated goal is ``uniting genetics and evolutionary biology with ecology''\cite{Haller2019}. The primary motivation for its creation was the observation that an efficiency/flexibility trade-off is unavoidable when designing a population genetics simulator; programs relying on coalescent approximations often sacrifice the flexibility needed to model complex biological and population phenomena. Nevertheless, \pkg{SLiM} manages to bridge some of the performance gap while providing an impressive array of features. Indeed, its latest major release can even be used to simulate a complete human genome \cite{Haller2025}.

In addition to chromosomal crossover and gene conversion, selection, complex scenarios of demography and population substructure, selective sweeps, epistasis, continuous spatial models, multiple species, and multiple chromosomes are supported. A distinct characteristic of \pkg{SLiM} is that many of these features are not hard-coded but instead left for users to implement. This is accomplished through a domain-specific language (DSL) called \proglang{Eidos}, specifically designed for this purpose. \proglang{Eidos} is admittedly ``very simple and unoriginal,'' with its syntax and semantics heavily inspired by \proglang{C} and \proglang{R}. Consequently, it is relatively easy to learn, even for users with little to no programming experience. One of the uses of \pkg{Eidos} is to specify initial conditions for a simulation run, and in this respect, it is functionally equivalent to a well-designed, text file-based parameter specification system. What sets it apart is the possibility to finely tune the simulation workflow via a callback mechanism, allowing user-defined functions to be called at specific stages. This enables users to implement functionalities not previously envisioned by \pkg{SLiM}'s authors without having to modify and recompile the program's source code.

\pkg{SLiM} is compatible with non-WF forward-in-time agent-based models, which are characterized by an emphasis on individual behavior and, as a result, increased biological realism. Much of the support for this class of models comes from explicitly modeling individual ages, and from an enriched API and additional callbacks. The former allows for overlapping generations and varying population sizes by modeling fitness as survival probabilities. The latter gives users fine control over migration and reproduction events.

The price for \pkg{SLiM}'s flexibility is having to learn a new programming language, albeit a relatively easy one. However, this is partly offset by a comprehensive collection of ``recipes'' for common scenarios available in the manual. Additionally, a graphical interface called \pkg{SLiMgui} is provided to assist in the development of \proglang{Eidos} scripts. Many of the functionalities of an integrated development environment are available, including code completion and highlighting, as well as static syntax checking. Simulations can be observed in real time, with the possibility of live graphical representations.

Another option is to use \pkg{SLiM} through \proglang{R}, facilitated by the \pkg{slimr} \proglang{R} package\cite{Dinnage2023}, available on GitHub. Note that this package is different from \pkg{SlimR} available on CRAN.

Much of the performance of \pkg{SLiM} comes from using optimized data structures, including \code{treeSequence}s, whose implementation is discussed in \cite{Haller2019a}. \pkg{SLiM} is tightly integrated with the \pkg{tskit} ecosystem, which can be used for post-simulation analysis. Moreover, a neutral ``burn-in'' coalescent simulation from \pkg{msprime} can be prepended to a genealogy, an operation dubbed ``recapitation.'' Users therefore benefit from the best of both worlds: the realism of forward-in-time simulation and the performance of a coalescent approximation. Recapitation, among other operations, can be performed using a small \proglang{Python} package called \pkg{pyslim} available on PyPI.

\pkg{SLiM}'s documentation and the manual for the \proglang{Eidos} language are extensive. Both are available on the project's website. Due to its ability to output \code{treeSequence}s, the simulation results are readily converted to any format supported by \pkg{tskit}. Source code is available on the project's GitHub repository, along with a macOS installer and instructions for installation on Linux and Windows.

\subsubsection{\pkg{scrm} \cite{Staab2015}}
\classification{\cg}{\cg}{\cg}{\cg}{\proglang{C++}}{CLI}

\pkg{scrm}, the software, is an implementation of SCRM, an approximation of the standard coalescent with recombination model. It is presented as an improvement on SMC and SMC', both of which may be unsuitable under certain conditions \cite{Eriksson2009}. From a performance standpoint, reported simulation studies suggest that \pkg{scrm} is more efficient than either \pkg{MaCS} or \pkg{FastSimCoal} and about 50 times faster than ms. Part of this speed-up is related to the data structure. Indeed, \pkg{scrm} avoids the complexities associated with graphs by storing information in an efficient tree-based data structure. From an algorithmic standpoint, it shares multiple similarities with simulators of the \pkg{MaCS} family in general and \pkg{SC} in particular. Type 2 recoalescence events are generated by allowing recoalescence with non-ancestral branches and generating a new recombination event on the recoalescence branch with a certain probability. Both programs can approximate the full CWR distribution by disregarding branches separated from the current marginal tree by a distance above a user-defined threshold. In that regard, \pkg{scrm}'s justification for disregarding linkage passed a certain distance is well detailed. It is framed in terms of \emph{accumulated recombination rate}. Simply put, the rate at which recombination events occur on non-ancestral branches increases with its distance from the current marginal tree. A consequence of this observation is that recoalescence events with distant branches might as well be ignored, as they are unlikely to affect the history of the recoalescing tree. However, in practice, ignoring branches with a high recombination rate is equivalent to constraining events to occur on branches within a sliding window of ancestral material. Interestingly, \cite{Kelleher2016} reports that there does not seem to be a window size above which approximation of the CWR worsens, which is the case for \pkg{MaCS}.

\subsubsection{\pkg{Relate} \cite{Speidel2019}}\label{relate}
\classification{\xg}{\cg}{\cg}{\xg}{\proglang{C++}}{CLI}

\pkg{Relate} implement an ARG inference algorithm whose primary goal is to enable analysis of large samples on a genome scale. It can manage on the order of $10^4$ phased and polarized sequences of biallelic SNPs and allows for some level of error in the input data. Another feature of interest is the possibility of constraining ancestries by specifying ancient sequences \cite{Speidel2021}.

\pkg{Relate}'s algorithm generates a sequence of trees which is consistent with the input sample in certain limiting scenarios. Fundamentally, marginal tree reconstruction is dealt with as a hierarchical clustering problem. Clustering is performed according to a distance related to the difference in the number of mutations on each sequence. The distance in question is not a metric since it is not symmetric; this is intentional to take advantage of the polarized nature of the data. In fact, it is constructed from the concept of \emph{derived mutation}, which is the number of derived alleles carried by one of the haplotypes but not by the other. The number of derived mutations between each pair of haplotypes is computed from the posterior probabilities of a Li-Stephens model modified to account for polarization. The result is forwarded to the tree builder, the procedure tasked with performing the clustering. Ties are possible and dealt with heuristically. Importantly, branch length is not estimated at this stage.

Tree building is not performed at each SNP but rather when needed. The computationally demanding procedure is only performed when a marker cannot be \emph{mapped} to the current marginal tree. The mapping rules follow the ISM model approximately. Ideally, it should be possible to locate a single mutation event on the marginal tree in a way that separates derived sequences from those carrying the wild allele. In practice, some flexibility is allowed to account for the various errors that may affect the data.

Branch length estimation is performed via MCMC once topology reconstruction is complete. A coalescent prior is used and equivalent branches in different trees are reconciled before the beginning of the sampling procedure. As noted by the authors, this part of the algorithm is computationally demanding but allows for great flexibility.

Functionalities of \pkg{Relate} can be extended via add-ons. These add-ons provide capabilities such as effective population size, mutation rates, and branch lengths estimation/reestimation, selection detection, tree extraction according to user-specified criteria, and plotting. These add-ons are included in the standard \pkg{Relate} distribution as standalone scripts.

Conveniently, shared memory parallelization can be achieved via the included \code{RelateParallel.sh} script, which takes the same arguments as the main program, with the addition of the \code{\textendash\textendash threads n} switch to specify the number of threads to use. Distributed memory parallelism is also possible, but requires a more involved setup. A script to run simulations on a ``Sun Grid Engine'' managed cluster is provided, which, while being written for a deprecated system, should be compatible with the Open Cluster Scheduler.

Input haplotypes are to be provided in the form of haps/legend/sample files, which is the format used by \pkg{SHAPEIT} \cite{Delaneau2011}. When necessary, recombination maps can be provided via a custom format detailed in the online documentation.

\subsubsection{\pkg{tsinfer} \cite{Kelleher2019}}\label{tsinfer}
\classification{\xg}{\cg}{\cg}{\xg}{\proglang{C}}{\proglang{Python}}

\pkg{tsinfer} takes the problem of ancestry inference backward. It starts by inferring ancestral haplotypes and then works its way back to the sample. Its algorithm is divided into three steps, the first being to determine the age of the derived allele for each marker. This is done using a simple heuristic relying on derived allele frequencies. The term ``age'' is used loosely as only the relative temporal ordering is estimated.

The ancestral haplotypes are inferred in the second step. Estimation is informed by the results of the preceding stage. Again, the process is entirely heuristic and very efficient, relying on a consensus algorithm rather than a genetic model. For each focal site, a set of descendants is constructed assuming the ISM. This set contains every haplotype with the derived allele. The state of nearby sites is determined based on consensus between haplotypes. Disagreement with the consensus is interpreted as indicative of recombination. Divergent haplotypes are removed from the set of descendants as inference propagates left and right. Inference stops once the cardinality of the sets goes below a certain threshold. This process does not generally span the full length of the haplotypes; the remaining markers are deemed non-ancestral.

The final step of the algorithm is to associate each haplotype with a set of ancestors in order to produce an ARG in the form of a sequence of marginal trees. Ancestors are computed along haplotypes by maximum likelihood under the LSM. One peculiarity of \pkg{tsinfer} compared to other programs using the LSM, such as \pkg{Threads} or \pkg{Relate}, is that the reference panel is made up of ancestral rather than sampled haplotypes. This is problematic when the number of markers $m$ is large, since the size of the panel is on a similar scale, leading to a $\mathcal O(m^2)$ time complexity. Moreover, the reference panel is not static through this step. To circumvent these issues, the LSM is formulated in terms of Tree Sequences (see \cref{sec:msprime}) in a way that makes computation and storage of the likelihood value unnecessary for certain ancestral haplotypes. To actually perform the inference, \pkg{tsinfer} implements a custom version of the Viterbi algorithm that is aware of the relationships between the elements of the reference panel. Due to the ancestry's topology, many sequences have the same likelihood as their parents. These haplotypes are simply marked with a reserved value instead of undergoing likelihood computation. Without surprise, the end result is output as a \code{treeSequence}.

Despite its reliance on numerous heuristics, reported simulation studies suggest good statistical performance in practice. \pkg{tsinfer} is distributed as a Python package and available on PyPI.

\subsubsection{\pkg{SARGE} \cite{Schaefer2021}}\label{sarge}
\classification{\xg}{\xs}{\xs}{\xg}{\proglang{C++}}{CLI}

\pkg{SARGE} is an ARG inference program which implements an heuristic that does not compromise when it comes to performance. It implements a greedy algorithm built upon the four gametes test. The author fully acknowledges the tendency of their program to systematically underestimate the number of recombination events and warns users against using it for tasks sensitive to such bias. Moreover, \pkg{SARGE}'s main focus is topological. Its approach to branch length estimation is a heuristic involving the locations and positions of mutation events. Despite this, simulation studies presented in the paper suggest that estimates are devoid of systematic biases. Additionally, recombination breakpoints are not simulated.

\pkg{SARGE} revolves around a simple procedure designed to determine a sequence of recombination events between two phylogenetic trees, made possible by a representation of binary trees as hierarchical clades of haplotypes. The main idea is to reorganize a subset of the clades of the first tree in a way that yields the second one. Since multiple choices are possible in general, the decision is guided by a parsimony heuristic: branch movements (i.e., SPR operation) explaining the maximum number of discordant clades are selected. Breakpoints are positioned halfway between the rightmost position associated with the left tree and the leftmost position associated with the right one. Discordance is defined as a failure of the four gamete test. As remarked earlier (\cref{gamarg}), applying this test to every pair of markers is not feasible in practice. \pkg{SARGE} overcomes this difficulty by disregarding markers positioned outside of a moving \emph{propagation window}, an approach similar to that of other algorithms such as \pkg{MaCS} or \pkg{scrm}.

\pkg{SARGE} uses a custom format for input files but is distributed with utilities for converting \code{VCF} files as well as the output of a run of the \pkg{ms} simulator into it. Other notable utilities include \proglang{Python} scripts for Newick-formatted trees visualization, propagation window determination, and gap determination. The latter can be used to split haplotypes into subsequences suitable for parallel analysis, which can be performed by an included \proglang{bash} script relying on \pkg{GNU Parallel}.

A peculiarity of \pkg{SARGE} is that the sample size is a compile-time constant. This, combined with the fact that it is written in \proglang{C++} and is hence ahead-of-time compiled, may result in significant overhead, temporal or spatial, for the practitioner dealing with samples of variable sizes on a regular basis. \pkg{SARGE} source code as well as a detailed readme are available on GitHub.

\subsubsection{\pkg{ARGinfer} \cite{Mahmoudi2022}}\label{arginfer}
\classification{\cg}{\cg}{\cg}{\xg}{\proglang{Python}}{\proglang{Python*}}

The impressive computational performance of software such as \pkg{Relate} and \pkg{tsinfer} allows them to handle large samples, but this comes at the price of various approximations. On the other hand, programs implementing an MCMC scheme such as \pkg{ARGweaver} and \pkg{Arbores} enjoy good statistical properties but are limited to sample sizes a few orders of magnitude smaller than those of their recent heuristic counterparts. Importantly, these are based on the SMC model instead of the full CWR and cannot generate genealogies containing trapped non-ancestral material. Like these two programs, \pkg{ARGinfer} implements probabilistic ARG inference via an MCMC algorithm. However, it does so by targeting the CWR distribution instead of the SMC, hopefully improving accuracy in the process. From a computational performance perspective, ancestries are stored in a custom data structure called \emph{Augmented Tree Sequence} (ATS), a variation on \pkg{tskit}'s Tree Sequence designed for ARG inference.

\pkg{ARGinfer}'s assumes a coalescent prior and uses a likelihood that takes branch lengths, genomic intervals, and number of mutations carried into account. The initial ARG is inferred using a heuristic algorithm inspired by ARG4WG (\cref{arg4wg}). The proposal distribution is constructed from six different types of moves the first four being relatively straightforward: an SPR operation, the addition/removal of a recombination, and the resampling of a recombination breakpoint. These are readily implemented but are chosen far less often in a typical run than the other two moves, which have superior mixing capabilities. These are the Kuhner move and the resampling of event times. A Kuhner move is similar to an SPR operation but follows the full CWR distribution. Specifically, the pruned edge can undergo a sequence of recombination and coalescence events before completely recoalescing with the ARG.

An ATS record is a 6-tuple $(s, p, c, b, m, t)$ where $t$ and $c$ are defined as in a coalescence record for a Tree Sequence (\cref{sec:msprime}). $b$ and $m$ contain the eventual breakpoints and mutation events associated with the branch. The entries $l$ and $r$ of a coalescence record are replaced by $s$, a sequence of disjoint regions of the form $[l_1, r_1], \ldots, [l_k, r_k]$. $p$ contains the parents of the edge. If the event at the bottom of the branch is a coalescence, $c$ is a pair, just as it always is in the case of coalescence records. However, a major difference between these and ATS records is the latter's capacity to encode recombination events directly. When one such event is at the bottom of a branch, the associated $c$ is a singleton. The reverse is true for $p$: it is a singleton when the top event is a coalescence and a pair when it is a recombination. Topologically, an ATS is a simple graph. Reportedly, storing mutation events along with edges helps speed up computations by allowing likelihood evaluation without site-by-site sequence comparisons. Despite this, numerical experiments on simulated data suggest that \pkg{ARGinfer} takes longer than \pkg{ARGweaver} to achieve convergence. This is especially true when the mutation rate is small compared to the recombination rate. According to the authors, this discrepancy is due in part to the poor mixing provided by \pkg{ARGinfer}'s proposal distribution. On the flip side, parameter estimation performances seem approximately at least as good as \pkg{ARGweaver}'s and even superior for recombination rates, TMRCA, and allele age (i.e. mutation latitude).

In addition to a \proglang{Python} API, a CLI interface is available. Both interfaces are documented in the package documentation hosted on Read the Docs. Source code is available on GitHub, and the package is distributed through PyPI.

\subsubsection{\pkg{Espalier} \cite{Rasmussen2023}}\label{espalier}
\classification{\cg}{\xs}{\xs}{\xg}{\proglang{Python}}{\proglang{Python*}}

Admittedly, the primary target of the botanically-named \pkg{Espalier} is not human genomic data. However, we feel that the originality of its approach to inference, coupled with the fact that it can be applied to human data regardless of its creators' intent, warrants its inclusion in this review.

On the surface, the most striking characteristic of \pkg{Espalier} is its ability to generate recurrent mutations. In this respect, it can be considered the model-based counterpart of \cite{Ignatieva2021}. In addition, for reasons that will become clear in subsequent paragraphs, \pkg{Espalier} works best in scenarios where the mutation rate ``significantly exceeds'' the recombination rate. Overall, this makes it well-suited for ancestry inference of rapidly mutating organisms. Indeed, \cite{Rasmussen2023} report a simulation study performed on a sample of genomes of potyviruses, a plant RNA virus with a high mutation rate.

The algorithm revolves around a preexisting concept intimately related to SPR, known as a maximum agreement forest (MAF) \cite{Hein1996}. Removing edges from a tree creates a set of disconnected trees called a \emph{forest}. An \emph{agreement forest} for two trees is any forest obtainable in this way from both trees. Given two trees with $n$ leaves, finding an agreement forest is trivial (by removing $n - 1$ edges, for instance). However, computing a forest minimizing the number of cuts is, in general, provably NP-hard. Such a forest is a MAF. Consequently, \pkg{Espalier} relies on a custom approximate algorithm named the \emph{4-cut MAF algorithm}, which is a modified version of the \emph{3-cut MAF algorithm}.

Somewhat similarly to \cite{Heine2018}, the sample of genetic material for which an ancestry is to be inferred needs to be divided into regions. Each of these regions should have a low probability of recombination events occurring within them and is determined by the user. \pkg{Espalier}'s algorithm demands that a coalescent tree be associated with each region. These can either be user specified or constructed by the program. According to the authors, the accuracy of these initial trees is a relatively minor concern, as their algorithm leverages correlations between neighboring trees to improve them. In any case, this concludes the first stage of the algorithm. In the following stage, pairs of trees are iteratively reconciled via a branch-and-bounds type algorithm. Given two trees $T_1$ and $T_2$ and an (approximate) MAF $V = V_1, \ldots, V_m$ obtained from the 4-cut algorithm, a tree is reconstructed by iteratively regrafting $V_k, k = 2, \ldots, m$ to the trees generated in the previous steps of the procedure. At step $k$, $2^k$ trees are created by reattaching $V_{k + 1}$ to its parent edge in either $T_1$ or $T_2$. In practice, only trees for which a likelihood ratio is above a certain bound are considered, which contributes to making the algorithm tractable. Once the reconciliation procedure is over, a sequence of trees is selected by maximizing the joint likelihood of marginal trees as provided by the SMC distribution. This task is framed as a HMM problem and carried out efficiently via the Viterbi algorithm. Ultimately, latitudes are assigned to vertices via a tailor-made algorithm and recombination events are imputed with the help of previously computed MAFs. The final result is stored as a \pkg{tskit} object. As an aside, the authors provide a method for constructing multiple ancestries from the same sample.

The simulation experiments reported in the paper cover a broad spectrum of scenarios and are discussed at length. The reported running time of \pkg{Espalier} is at least an order of magnitude faster than the convergence time of \pkg{ARGweaver}. Despite this, it remains slow in comparison to other ARG inference software. Its author suggests that the pure \proglang{Python} implementation of the 4-cut MAF algorithm may leave performance improvements on the table.

\pkg{Espalier} is distributed as a \proglang{Python} package on PyPI and is documented on Read the Docs. Its source code is publicly available on GitHub. A CLI is available in addition to the \proglang{Python} interface.

\section{Conclusion}
We have reviewed thirty-two ancestral recombination graph simulation and inference software. Our findings are summarized in \cref{table:summary}. Among them, about two-thirds are designed to infer ancestries, while the others implement simulation algorithms. The most striking features of our dataset are as follows:
\begin{itemize}
\item Every heuristic-based program is an inference program. Equivalently, ARG simulators are model-based.
\item Almost half of the algorithms described disregard either type B coalescence events or type 2 recombination events. Most of those are inference algorithms.
\item \proglang{C} and \proglang{C++} are by far the most common programming languages.
\item Most programs have a single interface, and that interface is a command line.
\end{itemize}

ARG inference programs' heavy dependence on heuristics reflects the challenges involved in reconstructing genealogies. Formulating a statistical model for sample-consistent ARGs is difficult. Designing scalable inference algorithms can prove even more challenging, and implementing them correctly is not a task to be underestimated. These observations help explain why, to a large extent, all is fair in ARG inference. Consistent ancestries are valuable, and many are prepared to sacrifice some statistical correctness to obtain them, more so than in other inference contexts. Similar observations can be made regarding the widespread exclusion of certain categories of events by this class of software. In this regard, the variable level of approximation offered by \pkg{SC-sample} is an interesting feature, allowing users to finely tune the trade-off between performance and realism to suit their needs.

The ubiquitous use of \proglang{C} and \proglang{C++} for implementation is hardly surprising. Both are mature and well-established programming languages with a largely justified reputation for delivering robust performance. Hence, they are natural choices for tackling the computational challenges inherent to ARG simulation and inference. The main drawback of these languages is perhaps the awkwardness involved in integrating them into a modern workflow. There are no formal packaging standards for these languages, and no standardized way of sharing code, at least nothing with a scope comparable to that of \proglang{Python} for instance. Even though platforms such as GitHub help, distribution remains challenging. This is compounded by the fairly involved process of compiling \proglang{C}/\proglang{C++} source code. On that matter, the ahead-of-time compilation required by these languages makes designing interactive software a highly non-trivial task. As a matter of fact, the vast majority of the programs reviewed are limited to a command-line interface and plain-text input/output. This is not inherently bad: CLIs are ubiquitous, and plain text is a famously universal interface. That being said, universality is not a replacement for convenience. In many contexts, the ability to interact with a library via a well-designed API is invaluable in streamlining workflows. Arguably, a polished interface can consolidate adoption among practitioners. In that regard, \pkg{msprime} is one of a kind. It is by far the most widespread simulator. While its performance and flexibility are certainly a major contributing factor to its popularity, the choice of its designers to make it available as a rich \proglang{Python} library is undoubtedly beloved by its numerous users.

Despite this success story, let us not forget that developing solutions that are both fast and user-friendly is hard: \pkg{msprime} is a major undertaking. For one thing, while it is mostly written in \proglang{Python}, \proglang{C} accounts for almost 40\% of the codebase. Interfacing both languages is not trivial, and coding in \proglang{C} is certainly harder than doing so in a high-level language like \proglang{Python}. There is hope that the recent popularity of recent languages such as \proglang{Julia}, which promise to simplify scientific programming by, among other things, solving the two-language problem, may speed up development and increase the quality of future software.

\begin{table}
  \centering

  \renewcommand{\arraystretch}{1.1}
  \begin{tabular}{*{3}{l}*{6}{c}}
    \textbf{Name} & \textbf{Year} & \textbf{Reference} & \textbf{Model} & \textbf{Coal. (B)} & \textbf{Rec. (2)} & \textbf{Simulated} & \textbf{Language} & \textbf{Interface}\\
    \hline

    \\\multicolumn{9}{c}{\textbf{ms Family}}\\
    \pkg{ms} & 2002 & \cite{Hudson2002} & \cg & \cg & \cg & \cg & \proglang{C} & CLI\\
    \pkg{msms} & 2010 & \cite{Ewing2010} & \cg & \cg & \cg & \cg & \proglang{Java} & CLI\\
    \pkg{cosi2} & 2014 & \cite{Shlyakhter2014} & \cg & \cg & \cg & \cg & \proglang{C++} & text\\
    \pkg{discoal} & 2016 & \cite{Kern2016} & \cg & \cg & \cg & \cg & \proglang{C} & CLI\\
    \pkg{msprime} & 2021 & \cite{Baumdicker2021} & \cg & \cg & \cg & \cg & \proglang{C} & \proglang{Python}*\\
    \\\multicolumn{9}{c}{\textbf{SIMCOAL Family}}\\
    \pkg{SIMCOAL2} & 2004 & \cite{Laval2004} & \cg & \cg & \cg & \cg & \proglang{C++} & CLI\\
    \pkg{fastsimcoal} & 2011 & \cite{Excoffier2011} & \cg & \xs & \xs & \cg & \proglang{C++} & CLI\\
    \\\multicolumn{9}{c}{\textbf{SHRUB Family}}\\
    \pkg{SHRUB} & 2005 & \cite{Song2005a} & \xg & \cg & \cg & \xg & \qm & \qm\\
    \pkg{beagle} & 2005 & \cite{Lyngsoe2005} & \xg & \cg & \cg & \xg & \proglang{C} & CLI\\
    \pkg{KwARG} & 2021 & \cite{Ignatieva2021} & \xg & \cg & \cg & \xg & \proglang{C} & CLI\\
    \\\multicolumn{9}{c}{\textbf{Margarita Family}}\\
    \pkg{Margarita} & 2006 & \cite{Minichiello2006} & \xg & \xs & \xs & \xg & \proglang{Java} & CLI\\
    \pkg{ARG4WG} & 2017 & \cite{Nguyen2017} & \xg & \xs & \xs & \xg & \proglang{C++} & CLI\\
    \pkg{GAMARG} & 2019 & \cite{Thao2019} & \xg & \xs & \xs & \xg & \proglang{C++} & CLI\\
    \\\multicolumn{9}{c}{\textbf{MaCS Family}}\\
    \pkg{MaCS} & 2008 & \cite{Chen2008} & \cg & \cg & \xs & \cg & \proglang{C++} & CLI\\
    \pkg{SC} & 2014 & \cite{Wang2014} & \cg & \cg & \cg & \cg & \proglang{C++} & CLI\\
    \pkg{SC-sample} & 2014 & \cite{Wang2014} & \cg & \cg & \cg & \xg & \proglang{C++} & CLI\\
    \\\multicolumn{9}{c}{\textbf{Tree Scan Family}}\\
    \pkg{RENT} & 2011 & \cite{Wu2011} & \xg & \cg & \cg & \xg & \proglang{C++/Java} & CLI\\
    \pkg{ARBORES} & 2018 & \cite{Heine2018} & \cg & \xs & \xs & \xg & \proglang{C} & CLI\\
    \\\multicolumn{9}{c}{\textbf{ARGWeaver Family}}\\
    \pkg{ARGweaver} & 2014 & \cite{Rasmussen2014} & \cg & \xs & \xs & \xg & \proglang{C++} & CLI\\
    \pkg{ARG-Needle} & 2023 & \cite{Zhang2023} & \xg & \xs & \xs & \xg & \proglang{C++} & CLI*\\
    \pkg{Threads} & 2024 & \cite{Gunnarsson2024} & \cg & \xs & \xs & \xg & \proglang{C++} & CLI\\
    \pkg{SINGER} & 2025 & \cite{Deng2025} & \cg & \xs & \xs & \xg & \proglang{C++} & CLI\\
    \\\multicolumn{9}{c}{\textbf{Others}}\\
    \pkg{SNPsim} & 2003 & \cite{Posada2003} & \cg & \cg & \xs & \cg & \proglang{C} & CLI\\
    \pkg{FastCoal} & 2006 & \cite{Marjoram2006} & \cg & \xs & \xs & \cg & \proglang{C++} & \qm\\
    \pkg{GENOME} & 2007 & \cite{Liang2007} & \cg & \cg & \cg & \cg & \proglang{C++} & CLI\\
    \pkg{TMARG} & 2008 & \cite{Wu2008} & \xg & \cg & \cg & \xg & \proglang{C++} & \qm\\
    \pkg{SLiM} & 2013 & \cite{Messer2013} & \cg & \cg & \cg & \cg & \proglang{C/C++} & DSL*\\
    \pkg{scrm} & 2015 & \cite{Staab2015} & \cg & \cg & \cg & \cg & \proglang{C++} & CLI\\
    \pkg{Relate} & 2019 & \cite{Speidel2019} & \xg & \cg & \cg & \xg & \proglang{C++} & CLI\\
    \pkg{tsinfer} & 2019 & \cite{Kelleher2019} & \xg & \cg & \cg & \xg & \proglang{C} & \proglang{Python}\\
    \pkg{SARGE} & 2021 & \cite{Schaefer2021} & \xg & \xs & \xs & \xg & \proglang{C++} & CLI\\
    \pkg{ARGinfer} & 2022 & \cite{Mahmoudi2022} & \cg & \cg & \cg & \xg & \proglang{Python} & \proglang{Python*}\\
    \pkg{Espalier} & 2023 & \cite{Rasmussen2023} & \cg & \xs & \xs & \xg & \proglang{Python} & \proglang{Python*}
  \end{tabular}
  \caption{Summary of ARGs simulation and inference software. \hl{Checkmarks (\cg or \cs) and cross marks (\xg or \xs) indicate the presence or absence of a property, respectively. A green/red checkmark indicates desirable/undesirable features. The opposite is true for crosses. Up-to-date links to source code and documentation are available at} \url{https://patrickfournier.ca/publications/arg-software-review/graph}}\label{table:summary}
\end{table}

\bibliographystyle{unsrtnat}
\bibliography{refs}
\end{document}